%% file: main.tex
\documentclass[journal,letter]{IEEEtran}
\usepackage{amssymb}
\usepackage{amsmath}
\usepackage[table]{xcolor}
\usepackage{graphicx}
\usepackage{booktabs,tabularx}
\usepackage{multirow,hyperref}
\usepackage{ragged2e}
\usepackage{array}
\usepackage{tikz}
\usepackage{svg}
\usepackage{hyperref}
\usepackage{pgfplots}
\pgfplotsset{compat=1.18}
\usepackage{pifont}
\usepackage[edges]{forest}
\usepackage{xspace}
\usepackage[normalem]{ulem}
\usepackage{titlesec}
\usepackage{ltxtable}
\usepackage{cleveref}

\newcommand{\lc}[1]{}

\usetikzlibrary{decorations.pathreplacing,calligraphy}
\usetikzlibrary{arrows.meta,shapes,positioning,shadows,trees}
\usetikzlibrary{shadows.blur}

\newcommand{\JS} {}

\newcommand{\MIA}{MedIA\xspace}
\newcommand{\etal}{\textit{et~al}.\xspace}
\newcommand{\ie}{\textit{i}.\textit{e}.\xspace}
\newcommand{\eg}{\textit{e}.\textit{g}.\xspace}
\newcommand{\etc}{\textit{etc}.\xspace}
\newcommand{\REV} {}

\titleformat{\paragraph}{\normalfont\normalsize\itshape}{\theparagraph}{1em}{}
\titlespacing*{\paragraph}{0pt}{1ex}{0.1ex}
\titlespacing*{\subsection}{0pt}{1.5ex}{0.5ex}
\titlespacing*{\subsubsection}{0pt}{1ex}{0.1ex}

\newcommand{\parggg}[1]{\phantomsection\textbf{#1}}

\newcommand{\secref}[1]{{$\S$\ref{#1}}}
\newcommand{\pgref}[1]{{(p.~\pageref{#1})}}

\newcommand{\hyperfootref}[1]{%
    \hyperref[#1]{\footref{#1}}%
}

\newcolumntype{L}[1]{>{\raggedright\let\newline\\\arraybackslash\hspace{0pt}}m{#1}}
\newcolumntype{C}[1]{>{\centering\let\newline\\\arraybackslash\hspace{0pt}}m{#1}}

\newcommand{\cmrule}{\arrayrulecolor{black}\cmidrule(lr){2-8}\arrayrulecolor{black}}
\newcommand{\ccmrule}{\arrayrulecolor{lightgray!20}\cmidrule(lr){3-8}\arrayrulecolor{black}}
\newcommand{\ccmruleaa}{\arrayrulecolor{lightgray!20}\cmidrule(lr){3-7}\arrayrulecolor{black}}

\newcommand{\cmruledata}{\arrayrulecolor{lightgray!20}\cmidrule(lr){4-5}\arrayrulecolor{black}}
\newcommand{\cmruledataa}{\arrayrulecolor{lightgray!50}\cmidrule(lr){3-5}\arrayrulecolor{black}}
\newcommand{\cmruledataaa}{\arrayrulecolor{lightgray!80}\cmidrule(lr){2-5}\arrayrulecolor{black}}

\newcommand{\cmrulemethods}{\arrayrulecolor{lightgray!20}\cmidrule(lr){3-5}\arrayrulecolor{black}}
\newcommand{\cmrulemethodss}{\arrayrulecolor{lightgray!50}\cmidrule(lr){2-5}\arrayrulecolor{black}}
\newcommand{\cmrulemethodsss}{\arrayrulecolor{lightgray!80}\cmidrule(lr){1-5}\arrayrulecolor{black}}

\newcommand{\cmruleeq}{\arrayrulecolor{lightgray!20}\midrule\arrayrulecolor{black}}

\newcommand{\mrow}[2]{\multirow{#1}{1.5cm}{\centering #2}}

\title{Domain Generalization for Medical Image Analysis: A Review}
\author{Jee Seok Yoon, Kwanseok Oh, Yooseung Shin, Maciej A. Mazurowski, 
        and Heung-Il Suk,~\IEEEmembership{Senior Member,~IEEE}
        
\IEEEcompsocitemizethanks{
\IEEEcompsocthanksitem J. Yoon is with the Department of Brain and Cognitive Engineering, Korea University, Seoul 02841, Republic of Korea (e-mail: wltjr1007@korea.ac.kr).
\IEEEcompsocthanksitem K. Oh is with the Department of Artificial Intelligence, Korea University, Seoul 02841, Republic of Korea (e-mail: ksohh@korea.ac.kr).
\IEEEcompsocthanksitem Y. Shin is with the Department of Artificial Intelligence, Korea University, Seoul 02841, Republic of Korea (e-mail: usxxng@korea.ac.kr).
\IEEEcompsocthanksitem M. Mazurowski is with the Departments of Radiology, Biostatistics and Bioinformatics, Electrical and Computer Engineering, and Computer Science, Duke University, Durham, NC 27705, USA (e-mail: maciej.mazurowski@duke.edu)
\IEEEcompsocthanksitem H.-I. Suk is with the Department of Artificial Intelligence and the Department of Brain and Cognitive Engineering, Korea University, Seoul 02841, Republic of Korea, and the corresponding author (e-mail: hisuk@korea.ac.kr).
}
}%

\begin{document}

\maketitle
\input{sections/abstract}

\input{sections/introduction}
\input{sections/background}

\input{sections/methods}

\input{sections/insights}
\input{sections/conclusion}

\section*{Acknowledgement}
\vspace{-0.2cm}
This work was supported by Institute of Information \& communications Technology Planning \& Evaluation (IITP) grant funded by the Korea government (MSIT) RS-2022-II220959 ((Part 2) Few-Shot Learning of Causal Inference in Vision and Language for Decision Making) and RS-2019-II190079 (Department of Artificial Intelligence (Korea University)).

\bibliographystyle{IEEEtran}
\bibliography{references}

\input{biography}

\input{sections/appendix}

\end{document}

%% file: sections/abstract.tex
\begin{abstract}
\justifying{
Medical image analysis (\MIA) has become an essential tool in medicine and healthcare, aiding in disease diagnosis, prognosis, and treatment planning, and recent successes in deep learning (DL) have made significant contributions to its advances.
However, deploying DL models for \MIA in real-world situations remains challenging due to their failure to generalize across the distributional gap between training and testing samples --- a problem known as \emph{domain shift}.
Researchers have dedicated their efforts to developing various DL methods to adapt and perform robustly on unknown and out-of-distribution data distributions.
This paper comprehensively reviews domain generalization studies specifically tailored for \MIA.
We provide a holistic view of how domain generalization techniques interact within the broader \MIA system, going beyond methodologies to consider the operational implications on the entire \MIA workflow.
Specifically, we categorize domain generalization methods into data-level, feature-level, model-level, and analysis-level methods. We show how those methods can be used in various stages of the \MIA workflow with DL equipped from data acquisition to model prediction and analysis.
Furthermore, we critically analyze the strengths and weaknesses of various methods, unveiling future research opportunities.
}
\end{abstract}
\begin{IEEEkeywords}
Domain generalization, medical image analysis, out-of-distribution, deep learning
\end{IEEEkeywords}

%% file: sections/introduction.tex
\section{Introduction}
Medical image analysis (\MIA) plays a critical role in modern healthcare, enabling accurate diagnosis and treatment planning for various diseases.
Over the past few decades, deep learning has demonstrated great success in automating various \MIA tasks such as disease diagnosis~\cite{8585141}, prognosis~\cite{mukherjee2020shallow}, and treatment planning~\cite{duan2022evaluating}.
These achievements have become feasible by the capability of deep learning algorithms to learn from vast amounts of data, identify patterns, and generate predictive models that aid in \MIA tasks.
Moreover, the availability of powerful computational resources has greatly expedited the process of training deeper, wider, and more complex models.
These have led to impressive performance in relatively well-controlled settings. However, many challenges in real-world scenarios remain.

\input{figures/tikz/overview_pipeline}

With homogeneous data {\REV distributions}, well-designed models perform on par with and often surpass their human counterparts in many applications.
However, their reliability and robustness can be compromised when presented with previously unseen, out-of-distribution, or heterogeneous data.
This highlights a common challenge in the field of \MIA: the limited capacity of models to generalize to unfamiliar data distributions. Changes in data distribution can result from variations in imaging equipment, protocols, or patient populations.
Domain generalization aims to overcome these challenges by developing models that can adapt to new, unseen domains without compromising performance.

\subsection{Domain Generalization for Medical Image Analysis}
Domain generalization has emerged as a crucial field in deep learning, particularly in applications where the ability to generalize across diverse domains is of importance.
Its significance is particularly high in the context of \MIA, where data is very heterogeneous.
To better understand the unique challenges of domain generalization for \MIA, it is important to consider the following factors:
\begin{itemize}
    \item \textbf{Image appearance variability}: Variability in medical imaging refers to differences and inconsistencies typically manifest during the data acquisition process~\cite{nan2022data}. {\REV This} variability may arise externally from using different modalities, protocols, scanner types, and patient populations across multiple healthcare facilities, while internal variability may also occur within a controlled setting (\eg, same scanner or healthcare facility) due to factors such as hardware aging, software parameter variations, and human error (\eg, human motion).
    \item \textbf{Complex and high-dimensional data}: Medical images are often high-dimensional and may contain multiple channels or sequences. Many of such datasets span from thousands of pixels to gigapixel~\cite{chen2022scaling} and from 2-dimensions to 5-dimensions~\cite{feng20185d}. This complexity makes it difficult to identify and extract domain-invariant features that can generalize well across different domains.
    \item \textbf{Challenging data acquisition, organization, and labeling}: Large-scale, diverse, and labeled datasets are difficult to obtain due to the cost of data acquisition, privacy concerns, data sharing restrictions, and the labor-intensive nature of manual annotation by medical experts. Furthermore, quality assurance is challenging as the medical image is prone to noise and artifacts, such as patient motion, scanner imperfections, and imaging artifacts from hardware or software limitations.
    \item \textbf{Model interpretability, safety, and privacy}: In \MIA, ensuring model interpretability, safety, and compliance with regulatory and ethical standards is crucial. Robustness against adversarial examples and to out-of-distribution samples is essential to prevent adverse effects on patient care. Additionally, privacy-preserving data sharing and collaboration in multi-center contexts add complexity to implementing domain generalization techniques.
\end{itemize}

\subsection{Our Contributions}
With these challenging factors in mind, this review provides a comprehensive review of domain generalization techniques specifically tailored to \MIA.
There already exist few review papers on domain generalization with a specific focus on \MIA, but these are limited to specific data domain and task, \ie, mammography-based mass detection~\cite{ghosh2022large}, electroencephalography (EEG)-based emotion assessment~\cite{apicella2022machine}, and computational pathology~\cite{jahanifar2023domain}.
Also, there are several review papers on related topics for \MIA, such as domain adaptation~\cite{guan2021domain, sarafraz2022domain} and harmonization~\cite{nan2022data}.
However, domain generalization presents unique challenges compared to these tasks.

Multiple survey papers have been published that offer a comprehensive understanding of domain generalization for general data domains and tasks, presenting broader perspectives~\cite{wang2022generalizing,zhou2022domain,shen2021towards,csurka2021unsupervised,yang2021generalized,cui2022out,ghassemi2022comprehensive} as well as focused approaches such as causal models~\cite{sheth2022domain}, graph models~\cite{li2022out}, and federated learning~\cite{li2023federated}.
While these surveys serve as a detailed reference for specific algorithms, techniques, and model architecture, they lack an in-depth exploration of the system-level implications of domain generalization on the overall workflow of \MIA.

Our review aims to provide a holistic view of how domain generalization techniques interact within the broader structure of a \MIA system.
We go beyond the methodological hierarchy presented in previous surveys and delve into the operational consequences of domain generalization on the entire \MIA workflow (see Fig.~\ref{fig:overview_pipeline}).
Our focus is on understanding how domain generalization can be seamlessly integrated into every step of the decision-making process, including but not limited to data acquisition, pre-processing, model prediction, and analysis.
To this end, we categorize domain generalization techniques into each step of the \MIA workflow, \ie, from data preparation to analysis. 

\subsection{Scope of Review}
\input{figures/tikz/paper_count}
Literature search and selection were conducted by researchers experienced in machine learning and medical image analysis.
We used the Google Scholar search engine with three different search strategies, resulting in a database of 1,621 papers.
First, we searched for all papers that cited the existing domain generalization review~\cite{wang2022generalizing,zhou2022domain,shen2021towards,sheth2022domain,li2022out,sarafraz2022domain,apicella2022machine}.
Second, we searched Google Scholar using the exact phrase ``domain generaliz(s)ation'' and medical-related keywords (``Medical, CT, ultrasound, MRI, PET, X-ray, histology, histopathology, pathology, fundus, dermoscopy, endoscopy, mammography'') and selected 1,000 papers sorted by relevancy.
Lastly, we searched for the top 1,000 papers with the terms ``unseen'', ``domain'', and medical-related keywords.
The eligibility criteria for papers to be included in this review are that they have conducted at least one experiment involving the use of medical images within the domain generalization problem settings (see $\S$\ref{sec:problem-definition}), regardless of their use of the term ``domain generalization'' in their paper (small number of papers instead use ``unseen'', ``out-of-distribution'', or their own terms).
Peer-reviewed published papers were prioritized, but non-peer-reviewed archive papers (\eg, arXiv, bioRxiv) were also included if they had been deemed particularly suitable for selection (\eg, highly relevant, highly significant, highly cited).

%% file: figures/tikz/overview_pipeline.tex
\begin{figure*}[t]
    \centering
    
    \begin{tikzpicture}
    \node[inner sep=0pt] (scanner) at (0,0)
    {\includegraphics[height=.11\textwidth]{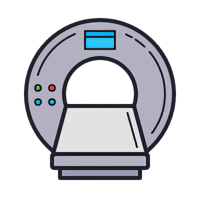}};

    \node[inner sep=0pt] (kspace) at (3,0)
    {\includegraphics[height=.1\textwidth]{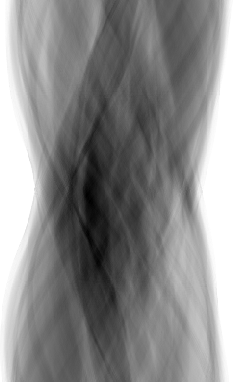}}; %
    
    \node[inner sep=0pt] (MRI) at (6,0)
    {\includegraphics[height=.1\textwidth]{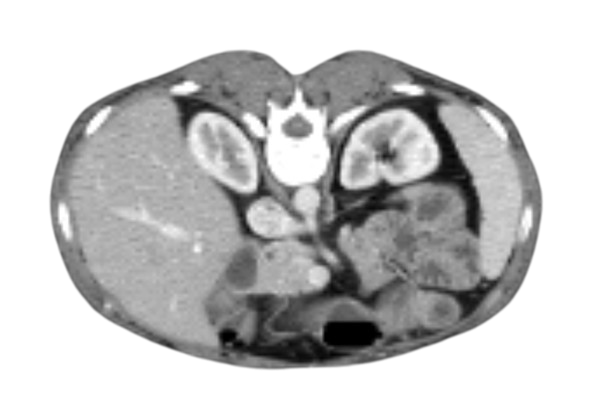}};

    \node[inner sep=0pt] (NN) at (9,0)
    {\includegraphics[height=.1\textwidth]{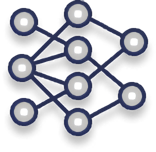}};

    \node[inner sep=0pt] (segmentation) at (12,0)
    {\includegraphics[height=.1\textwidth]{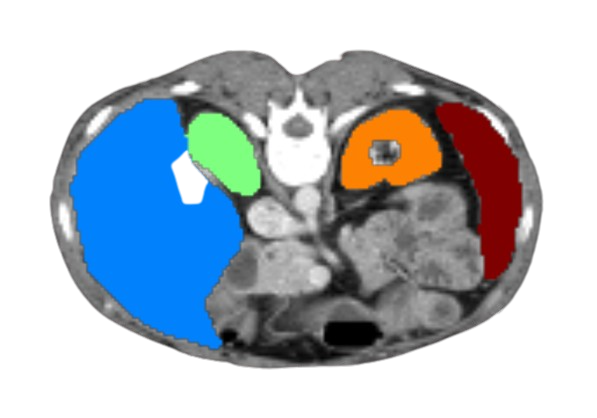}};
    
    \node[inner sep=0pt] (analysis) at (15.5,0)
    {\includegraphics[height=.1\textwidth]{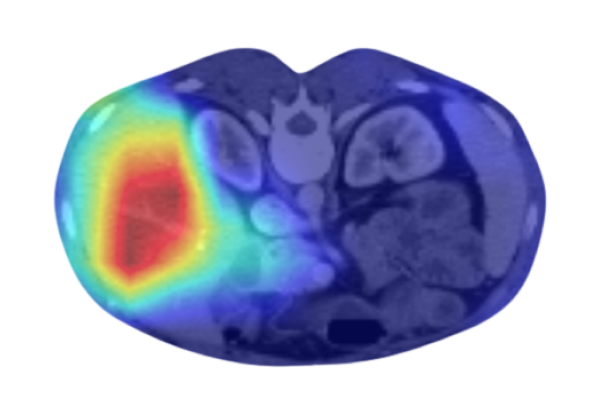}};

    \draw[->,thick] (scanner.east) -- (kspace.west)
    node[below, midway, text width=2.5cm,align=center] (data_acq_text) {\scriptsize Data Acq.};
    
    \draw[->,thick] (kspace.east) -- (MRI.west)
    node[below, midway, text width=2.5cm,align=center] (image_recon_text) {\scriptsize Image\\Recon.};

    \draw[->,thick] (MRI.east) -- (NN.west)
    node[below, midway, text width=2.5cm,align=center] (up_stream_text) {\scriptsize Upstream};
    
    \draw[->,thick] (NN.east) -- (segmentation.west)
    node[below, midway, text width=2.5cm,align=center] (down_stream_text) {\scriptsize Down-\\stream};
    
    \draw[->,thick] (segmentation.east) -- (analysis.west)
    node[below, midway, text width=2.5cm,align=center] (analysis_text) {\scriptsize Analysis};

    \draw[thick]
    (0, 1.22) -- (0, 1.6);
    \draw[thick,->]
    (0, 1.41) -- node[above, midway] {\footnotesize Data-level \secref{sec:data-level}} (6, 1.41);
    \draw[thick,->]
    (6, 1.41) -- node[above, midway] {\footnotesize Feature-level  \secref{sec:feature-level}} (9, 1.41);
    \draw[thick,->]
    (9, 1.41) -- node[above, midway] {\footnotesize Model-level  \secref{sec:model-level}} (12, 1.41);
    \draw[thick,->]
    (12, 1.41) -- node[above, midway] {\footnotesize Analysis-level  \secref{sec:analysis-level}} (15.5, 1.41)
    ;
    
    \node[inner sep=0pt, node distance=1.3cm] (data_acq_cite) [above of=data_acq_text] {\scriptsize Acq. standards~\cite{nan2022data}};
    \node[inner sep=0pt, node distance=1.475cm] (image_processing_cite) [above of=image_recon_text] {\scriptsize \quad Image processing \pgref{par:image-processing}};
    \node[inner sep=0pt, node distance=1.3cm, text width=2.5cm,align=center] (feat_alignment_cite) [above of=up_stream_text] {\scriptsize Alignment \pgref{sec:feature-alignment}};
    \node[inner sep=0pt, node distance=1.475cm] (learning_cite) [above of=down_stream_text] {\scriptsize \quad Learning strategy \pgref{sec:model-learning}};
    \node[inner sep=0pt, node distance=1.3cm] (trans_cite) [above of=analysis_text] {\scriptsize Transferability \pgref{sec:analysis-transferability}};

    \node[inner sep=0pt, node distance=1.3cm] (rawdata_cite) [below of=kspace] {\scriptsize Frequency-based \pgref{par:frequency}};
    \node[inner sep=0pt, node distance=1.3cm] (surrogate_cite) [below of=MRI] {\scriptsize Data augmentation \pgref{sec:data-augmentation}};
    \node[inner sep=0pt, node distance=1.3cm, text width=3cm,align=center] (opt_cite) [below of=NN] {\scriptsize Optimization \pgref{sec:model-optimization}};
    \node[inner sep=0pt, node distance=1.3cm] (causality_cite) [below of=segmentation] {\scriptsize Causality \pgref{sec:analysis-causality}};
    \node[inner sep=0pt, node distance=1.3cm] (interpret_cite) [below of=analysis] {\scriptsize Interpretable AI \pgref{sec:analysis-interpretable}};

    \draw[thick, dotted]
    (data_acq_text.north) -- (data_acq_cite.south)
    (image_recon_text.north) -- (image_processing_cite.south)
    (up_stream_text.north) -- (feat_alignment_cite.south)
    (down_stream_text.north) -- (learning_cite.south)
    (analysis_text.north) -- (trans_cite.south)
    ;

    \draw[thick, dotted]
    (kspace.south) -- (rawdata_cite.north)
    (MRI.south) -- (surrogate_cite.north)
    (NN.south) -- (opt_cite.north)
    (segmentation.south) -- (causality_cite.north)
    (analysis.south) -- (interpret_cite.north)
    ;

    \begin{scope}[shift={(-0.85,-1.2)}] %
        \draw[->, black, thick] (0,0) -- (0.42,0) node[right] {\scriptsize \MIA Workflow}; \draw[black, thick, dotted] (0,-0.4) -- (0.42,-0.4) node[right] {\scriptsize DG Techniques};
    \end{scope}

    \end{tikzpicture}
    
    \caption{Overview of the medical imaging analysis (\MIA) pipeline illustrating various stages and their associated domain generalization (DG) techniques. Stages include data acquisition, image reconstruction, upstream feature extraction, downstream task, and analysis. Each stage is associated with references to specific sections ($\S$), pages (p.), or external citations where the techniques are detailed.
    }
    \label{fig:overview_pipeline}
\end{figure*}

%% file: figures/tikz/paper_count.tex
\begin{figure}[h]\scriptsize
    \centering
    \begin{tikzpicture}
        \begin{axis}[
            width=1\linewidth, height=4cm,
            xticklabel style={rotate=330, /pgf/number format/set thousands separator={}},
            xtick={2013,...,2023},
            xtick distance=1,
            axis lines=left,
            tick label style={font=\footnotesize}
            ]
            
            \addplot table {
                Year Publications 
                2013 13
                2014 19
                2015 19
                2016 23
                2017 26
                2018 40
                2019 73
                2020 212
                2021 507
                2022 990
                2023 1510
            };
        \end{axis}
        \node[text width = 4cm, align = left] at (2.5,1.4){\tiny Keyword: ("domain generalization" OR "domain generalisation") AND (intitle:medical OR ct OR ultrasound OR mri OR pet OR xray OR histology OR histopathology OR pathology OR fundus OR dermoscopy OR endoscopy OR mammography)};
       
    \end{tikzpicture}
    \caption{Number of publications per year on the Google Scholar database.}
    \label{fig:paper-count}
\end{figure}

%% file: sections/background.tex
\section{Background}
\input{tables/notations}
\input{figures/tikz/settings}
\subsection{Problem Definition}\label{sec:problem-definition}
In this section, we formalize the problem of domain generalization (DG) by following the mathematical notations and formulations used in previous surveys~\cite{wang2022generalizing,li2023federated} (see Table~\ref{tab:notations} for the definition of mathematical notations).
Let $\mathcal{X}$ denote a nonempty input space and $\mathcal{Y}$ an output space (\eg, labels). A domain is composed of data that are sampled from a joint distribution of the input sample and output label $P_{XY}$. We denote a domain as $\mathcal{S} = \{(\mathbf{x}_j, y_j)\}^n_{j=1} \sim P_{XY}$, where $\mathbf{x} \in \mathcal{X}$, $y \in \mathcal{Y}$, and $n$ is the number of data pairs.

In DG, we are given $M$ training (source) domains $\mathcal{S}_{source}=\{\mathcal{S}^i \mid i=1,\cdots,M\}$, where $\mathcal{S}^i = \{(\mathbf{x}^i_j, y^i_j)\}^{n_i}_{j=1}$ denotes the $i$-th domain with $n_i$ data pairs.
The joint distributions between each pair of domains are different: $P^i_{XY} \ne P^{j}_{XY}, 1 \le i \ne j \le M$.
The goal of DG is to learn a robust and generalizable predictive function $h: \mathcal{X} \to \mathcal{Y}$ from the $M$ training domains to achieve a minimum prediction error on an $K$ \emph{unseen} test (target) domain $\mathcal{S}_{target}=\{\mathcal{S}^i \mid i=M+1,\cdots,M+K+1\}$ (\ie, $\mathcal{S}_{target}$ cannot be accessed in training).
In other words, the goal of DG is to minimize the generalization error:
\begin{equation}\label{eq:overall}
    \, \mathbb{E}_{(\mathbf{x},y) \in \mathcal{S}_{target}} [ \mathcal{L}(h(\mathbf{x}),y) ].
\end{equation}

\subsection{Settings of Domain Generalization}\label{sec:settings}
This subsection elucidates different settings for DG in the \MIA workflow, detailing the various configurations and challenges present in both the source and target domains during the implementation process (see Fig.~\ref{fig:settings}).
{\REV
A critical aspect of DG is overcoming the \emph{domain shift} --- differences between the source and target domains that hinder the model's ability to generalize. These gaps typically arise due to variations such as:
\begin{itemize}
\item Intensity Variations: Differences in image brightness or contrast, \eg, varying exposure levels in chest X-rays or staining intensities in histology images.
\item Resolution Differences: Variations in image detail, such as differences between high and low-resolution ultrasound devices or varying pixel densities in fundus photographs.
\item Noise Characteristics: Varying levels of image noise, \eg, speckle noise in ultrasound or grain in low-dose CT scans.
\item Artifact Patterns: Modality-specific artifacts, such as motion blur in MRI, beam hardening in CT, or light reflections in endoscopy images.
\item Anatomical Variations: Differences in patient populations, leading to variations in organ sizes or shapes across different datasets.
\item Label Distributions: Varying disease prevalence or severity between domains, affecting class balance in tasks like skin lesion classification from dermatology images.
\item Acquisition Protocols: Differences in imaging techniques, such as varying MRI sequences, CT reconstruction kernels, or staining protocols in histopathology.
\end{itemize}
As it is often challenging to attribute domain shifts to a single cause, we have systematically categorized these variations with into cross-site, cross-modality, cross-temporal, and cross-protocol domain shifts (see $\S$\ref{sec:settings_target}).
These shifts can be further classified into covariate and concept shifts (see $\S$\ref{sec:domain-shift}), providing a comprehensive framework for understanding and addressing domain shifts in \MIA.
}

\subsubsection{Settings for Source Domain} DG typically focuses on two settings regarding the number of source domains: \emph{multi-source DG} and \emph{single-source DG}~\cite{zhou2022domain}.
The multi-source setting assumes multiple distinct but relevant domains are available (\ie, $M > 1$). 
By leveraging the data from these domains, representations invariant to disparate marginal distributions are learned.
This is usually accomplished by minimizing the domain discrepancy among the source domains during the training process.
The single-source setting assumes training data is homogeneous (\ie, $M = 1$).
Therefore, this setting does not require domain labels during training.
Single-source DG tends to be more challenging than multi-source DG as it may not capture the full diversity of data distributions that exist across different domains.
Refer to Section~\ref{sec:source-limted-DG} for some extreme settings for domain generalization (such as, open-set DG, source-free DG, and unsupervised DG).

\subsubsection{Settings for Target Domain}\label{sec:settings_target} There are three settings for target domains that are unique to DG for \MIA as follows.
\begin{itemize}
    \item \textbf{Cross-site DG}: As the most prevalent form of DG for \MIA, the goal of cross-site DG is to develop models that can generalize well across different medical imaging datasets collected from multiple healthcare institutions. Cross-site DG helps in creating more robust models that can be deployed across different healthcare institutions without the need for extensive site-specific fine-tuning. 
    {\REV Cross-site DG research often define a domain or `site' as an individual healthcare facility, but some define it as an individual vendor, scanner, or even an individual patient when inter-patient variability is extremely large, which is a common setting in many EEG studies.}
    \item \textbf{Cross-sequence DG}: Medical imaging data often consists of multiple types of sequences or series, each capturing different aspects of the underlying anatomy or pathology. The most commonly used sequences are the \emph{cross-temporal} sequences and \emph{cross-protocol} sequences. Temporal sequences are images taken at different time points (\eg, before, during, and after treatment), while protocol sequences are images of the same modality with different acquisition protocols. For example, in magnetic resonance imaging (MRI), protocol sequences like T1-weighted, T2-weighted, and fluid attenuated inversion recovery (FLAIR) images provide different tissue contrasts and diagnostic information.
    \item \textbf{Cross-modal DG}: Medical imaging encompasses a wide range of modalities, such as MRI, computed tomography (CT), and X-ray. Each modality provides different types of information and is suited for specific clinical applications.
    This type of DG can involve training a model on data from one modality and testing its performance on data from a different, previously unseen modality.
\end{itemize}
    \subsubsection{Settings for Domain Shift}\label{sec:domain-shift}
    In the context of DG, domain shift can be categorized into \textit{covariate shfit} and \textit{concept shift}.
    \parggg{Covariate shift} happens when the data distribution between the source and target domains is different, but the functional relationship between the input and output (the ``concept'') remains the same.
    Given the source domain and a target domain, we have covariate shift when $P^{source}_X \ne P^{target}_X$ but $P^{source}_{Y|X} = P^{target}_{Y|X}$. Here, $P_X$ and $P_{Y|X}$ denote the marginal distribution of the input features and the conditional distribution of the output given the input, respectively.
    \parggg{Concept shift} occurs when the functional relationship between the input and output changes, \ie, $P^{source}_{Y|X} \ne P^{target}_{Y|X}$.

    To illustrate, consider two clinics that perform brain MRI scans on their patients. A covariate shift might be caused by differences in the MRI scanners, patient population, or other factors that affect the appearance of the brain scans. On the other hand, a concept shift might occur when the diagnostic criteria or the diseases of interest vary between the clinics.
    For example, one clinic might focus on diagnosing Alzheimer's disease, whereas another might concentrate on detecting brain tumors, in which case Alzheimer's disease might not be deemed significant.
    Additionally, the concept shift can manifest in the differences in diagnoses made by various medical professionals.
    This type of shift is closely associated with alterations in the assigned labels (label shift) or the interpretation of these labels (semantic shift).

    In the context of different DG settings, cross-site DG and cross-temporal DG could lead to covariate shift as the same concept (\eg, the presence or absence of a disease) may be associated with different input features (\eg, different patient populations) across different sites or at different times.
    In contrast, cross-protocol and cross-modal DG could potentially involve concept shifts.
    For example, a concept shift could occur when a model trained on MRI images, which highlights detailed information about soft tissues, struggles to correctly interpret CT scans that provide more detailed depictions of bone structures, essentially changing the underlying relationship between image features and the corresponding disease labels.

\input{figures/tikz/topology}
\subsection{Related Machine Learning Tasks}\label{sec:related}
\input{tables/related}

In this subsection, we discuss the relationship between DG and its related machine-learning tasks and clarify their differences. The main takeaway is that DG restricts its access to the target domain data, while other tasks have full or partial access to the target domain distribution. An overview of related tasks is in Table~\ref{tab:related}.

\begin{itemize}
    \item \textbf{Multi-task Learning (MTL)} aims to learn a single model that performs well on multiple related tasks. In the context of DG, MTL can be viewed as learning a predictive function $h$ that minimizes the combined risk over $M$ related tasks. The main difference between MTL and DG is that MTL aims to perform well on the same set of tasks that the model was trained on, while DG aims to generalize to unseen data distributions. 
    \item \textbf{Transfer Learning (TL)} aims to transfer the knowledge learned from one or more source domains to a different but related target domain. Both TL and DG deal with situations where the target distribution is different from the source distribution. However, in TL, the target domain is used during training (usually during fine-tuning), whereas in DG we assume no access to the target domain. 
    \item \textbf{Harmonization} aims to reduce non-biological heterogeneity caused by cohort bias (\eg, different scanner type or acquisition protocol). However, harmonization primarily focuses on cross-site datasets and does not necessarily impose restrictions on access to the target domain distribution. Most harmonization techniques are performed prior to model training mainly as a preprocessing technique.
    \item \textbf{Domain Adaptation (DA)} aims to tackle the domain shift problem encountered in new test environments. DA assumes the availability of labeled or unlabeled target data (\ie, \textbf{unsupervised DA, UDA}) for model adaptation. \textbf{Source-free DA (SFDA)} assumes source data is unavailable after pretraining a model (\eg, due to privacy reasons).
    \textbf{Zero-shot DA (ZDA)} limit its access to target domain data, but leverages auxiliary information related to the target domain. The primary distinction between UDA/SFDA/ZDA and DG lies in the (partial) access to target domain data during training.
    {\REV 
    \item \textbf{Zero-shot Learning (ZSL)} is closely related to out-of-distribution (OOD) generalization in that it aims to classify test samples with concept shift, but ZSL generally leverages auxiliary information, such as attribute descriptions, related to the target domain.
    \item \textbf{Test-time Adaptation (TTA)} deals with the domain shift problem as well. TTA differs from DA in that only a single or mini-batch of test data is used for model tuning, which is often done in an online manner.
    TTA and DG both share the constraint of not having access to the target domain during training. However, TTA requires an additional step of fine-tuning at test time, requiring a mini-batch of target data.
    \item \textbf{OOD Generalization} aims to detect the concept shift between in-distribution (ID) and OOD data. 
    While OOD and DG both assume no access to the target domain, the main difference between OOD and DG lies in that they focus on different domain shifts.
    Specifically, OOD mainly focuses on concept shift whereas DG considers both covariate and concept shift in their problem settings.}
\end{itemize}

%% file: tables/notations.tex
\begin{table}[ht]
\caption{Definition of mathematical notations, following~\cite{wang2022generalizing}.}
\vspace{-0.3cm}
\centering
\resizebox{0.5\textwidth}{!}{\begin{tabular}{llll}
\toprule
\textbf{Notation} & \textbf{Definition} & \textbf{Notation} & \textbf{Definition} \\
\toprule
$\mathcal{X}, \mathcal{Z}, \mathcal{Y}$ & Input, feature, output space & $\mathbf{x}, \mathbf{z}, y$ & Input, feature, output variables\\
$P_{XY}$ & Probability distribution  & $\mathcal{S}$ & Domain \\
$n_i$ & $i$-th domain data count & $M, K$ & Number of source, target domains \\
$\mathcal{L}(\cdot, \cdot)$ & Loss function & $h(\cdot)$ & Predictive function \\
$\mathcal{M}(\cdot)$ & Manipulation function & $f(\cdot)$ & Feature mapping function\\
$\mathcal{D}(\cdot, \cdot)$ & Dissimilarity function & &   \\
\bottomrule
\end{tabular}}
\label{tab:notations}
\end{table}

%% file: figures/tikz/settings.tex
\begin{figure*}[t]
    \centering
    \includegraphics[width=1\textwidth]{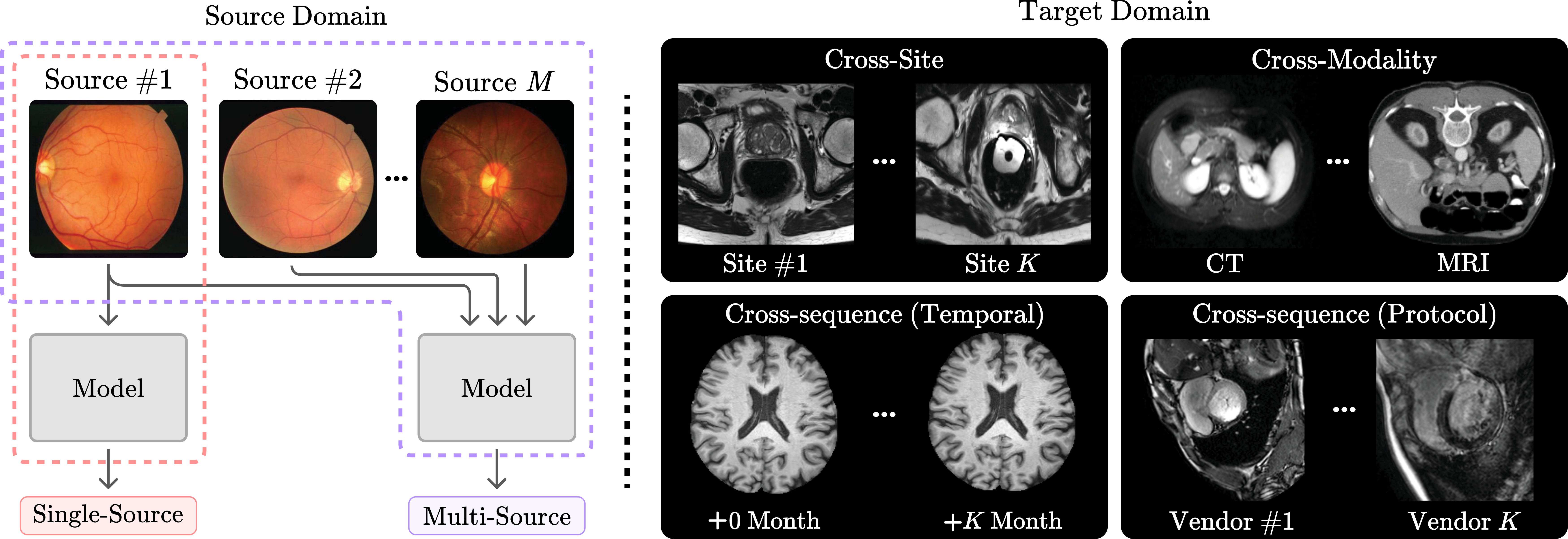}
    \caption{Settings of source and target domain for domain generalization.
    Source domain consists of $M$ domains, where $M=1$ refers to single-source and $M>1$ refers to multi-source settings.
    The target domain consists of $K$ domains.
    Cross-site, cross-sequence, and cross-modality define unique settings of the target domain for \MIA. Cross-site refers to generalization to different sites, \eg, different devices, and healthcare facilities. Cross-sequence refers to generalization to different sequences, \ie, different acquisition times or imaging protocols. Cross-modality refers to generalization to different modalities, \eg, CT to MRI.
    }
    \label{fig:settings}
\end{figure*}

%% file: figures/tikz/topology.tex
\begin{figure*}[thbp]
\centering
\begin{forest}
for tree = {
    draw=none, rounded corners,
    font=\footnotesize\linespread{0.84}\selectfont,
    text width=9em,
    text centered,
    calign=edge midpoint,
    edge = {-Stealth},
if level = 0{fill=teal!30, text width=15em}{},
if level>= 1{grow'=0,
             folder,
             folder indent=4mm,
             l sep=7mm,
             s sep=2mm,
             fill=teal!10,
             minimum height=2.5em,
    }{},
if level = 1{fill=teal!20, minimum height=0,}{},
            }
    [Domain Generalization for \MIA 
        [Data-level \secref{sec:data-level}
            [Manipulation \secref{sec:data-manipulation}]
            [Augmentation \secref{sec:data-augmentation}]
            [Problem-specific \secref{sec:data-problem-specific}]
        ]
        [Feature-level \secref{sec:feature-level}
            [Alignment \secref{sec:feature-alignment}]
            [Disentanglement \secref{sec:feature-disentanglement}]
            [Others \secref{sec:feature-others}]
        ]
        [Model-level \secref{sec:model-level}
            [Learning Strategies \secref{sec:model-learning}]
            [Model \mbox{Framework} \secref{sec:model-framework}]
            [Others \secref{sec:model-others}]
        ]
        [Analysis-level \secref{sec:analysis-level}
            [Interpretable AI \secref{sec:analysis-interpretable}
            [Transferability \pgref{sec:analysis-transferability}]]
            [Causality \secref{sec:analysis-causality}]
        ]
    ]
\end{forest}

    \caption{Hierarchical structure of the different aspects of domain generalization (DG) for medical image analysis (\MIA). This taxonomy divides the DG strategies into four primary levels: data-level, feature-level, model-level, and analysis-level, each encompassing distinct sub-strategies.}
    \label{fig:topology}
\end{figure*}

%% file: tables/related.tex
\newcommand*\emptycirc[1][1ex]{\tikz\draw[thick] (0,0) circle (#1);} 
\newcommand*\halfcirc[1][1ex]{%
  \begin{tikzpicture}
  \draw[fill] (0,0)-- (90:#1) arc (90:270:#1) -- cycle ;
  \draw[thick] (0,0) circle (#1);
  \end{tikzpicture}}
\newcommand*\fullcirc[1][1ex]{\tikz\fill (0,0) circle (#1);} 

\begin{table}[h]\scriptsize
\caption{Related machine learning tasks categorized by covariate and concept shift, and access to $\mathcal{S}_{target}$.}
\label{tab:related}
\centering
\begin{tabular}{cccc}
    \toprule
\textbf{Task} & \textbf{Covariate} & \textbf{Concept} & \textbf{$\mathcal{S}_{target}$} \\
    \toprule
Multi-Task Learning & &  & \fullcirc \\ \arrayrulecolor{lightgray!20}\midrule
Transfer Learning & \fullcirc & \fullcirc & \fullcirc \\\midrule
Harmonization & \fullcirc &  & \fullcirc \\\midrule
Domain Adaptation (DA) & \fullcirc &  & \fullcirc \\\midrule
Unsupervised/Zero-shot DA & \fullcirc &  & \halfcirc \\\midrule
Zero-shot Learning &  & \fullcirc & \halfcirc \\\midrule
Test-time Adaptation & \fullcirc &  & \halfcirc \\\midrule
Out-of-distribution &  & \fullcirc & \\\midrule
Domain Generalization & \fullcirc & \fullcirc &  \\\arrayrulecolor{black}
\bottomrule
\end{tabular}

\begin{center}
\footnotesize
\itshape
\fullcirc: Full access, \halfcirc: Partial access (\eg, auxiliary information, mini-batch).
\end{center}

\end{table}

%% file: sections/methods.tex
\section{Methods}
In this section, we review and explain a series of DG methods for medical imaging.
We employ a bottom-up approach and categorize the methods into data-level, feature-level, model-level, and analysis-level DG methods (see Fig.~\ref{fig:topology}).
Then, we explore some DG methods under extreme constraints (see $\S$\ref{sec:source-limted-DG}).
\begin{itemize}
    \item \textbf{Data-level generalization} methods focus on manipulating and generating input data to facilitate learning generalizable representations. 
    \item \textbf{Feature-level generalization} methods focus on extracting domain-invariant features from input images to improve the generalization performance of models. These methods often involve learning a shared feature representation across multiple domains by extracting domain-invariant features.
    \item \textbf{Model-level generalization} aims to improve DG in medical imaging by refining the learning process, model structure, or optimization techniques.
    \item \textbf{Analysis-level generalization} methods help users understand, explain, and interpret the decision-making process of machine learning models.
\end{itemize}
{\REV 
This categorization is inspired by the various stages of the \MIA pipeline, \ie, stages starting from data acquisition, image reconstruction, feature extraction, downstream task to analysis (see Fig.~\ref{fig:overview_pipeline}).
Our aim is to provide researchers and practitioners a holistic view of how different approaches can be combined and applied throughout the \MIA pipeline.

\begin{table}[ht]
\centering
{\REV
\caption{DG techniques applicable to multiple levels of the \MIA pipeline.}\label{tab:multiple-dg}
\begin{tabular}{c|ccc}
\toprule
 & Data-level & Feature-level & Model-level \\\midrule
Augmentation & \secref{sec:data-augmentation} & \secref{sec:feat_aug} &  \\
Adversarial & \secref{par:data-adversarial} & & \secref{par:model-adversarial}   \\
Contrastive &  & \secref{par:feature-contrastive} & \secref{par:model-ssl} \\ \bottomrule
\end{tabular}}
\end{table}

It is important to note that some DG techniques can be applied at multiple levels of the \MIA pipeline (see Table~\ref{tab:multiple-dg}). 
Augmentation at the data-level (\secref{sec:data-augmentation}) involves directly manipulating input images to increase data diversity, while feature-level augmentation (\secref{sec:feat_aug}) operates on extracted feature representations to expand the feature space. Adversarial techniques at the data-level (\secref{par:data-adversarial}) focus on generating challenging examples to improve robustness, whereas at the model-level (\secref{par:model-adversarial}) they aim to learn domain-invariant features through adversarial training. Contrastive learning at the feature-level (\secref{par:feature-contrastive}) encourages similar samples to have close representations while pushing dissimilar ones apart, while at the model-level (\secref{par:model-ssl}) it is used as a self-supervised pretraining strategy to learn generalizable representations. These multi-level applications of similar concepts highlight how DG techniques can be integrated throughout the pipeline, each tailored to address domain shift at different stages of processing.}

\input{sections/methods/data}
\input{sections/methods/feature}

\input{sections/methods/model}
\input{sections/methods/analysis}
\input{sections/methods/others}

%% file: sections/methods/data.tex
\subsection{Data-level Generalization}\label{sec:data-level}

\input{tables/data_level_dg}
The success of machine learning models often hinges on the training data's quality, quantity, and diversity.
As the qualitative acquisition of medical images is challenging and costly, data-level generalization methods present an efficient and straightforward approach to enhance a model's generalization capability.
These methods focus on manipulating and augmenting input data to increase the diversity and quantity of available samples, ultimately improving the model's adaptability to different domains.
Data-level generalization can be divided into two primary techniques: \textbf{Data manipulation}, which transforms existing data to expose the model to a broader range of samples, and \textbf{data augmentation}, which creates new samples to further expand the model's exposure to various data variations.
As these techniques are at the early stages of the \MIA workflow, \eg, data acquisition and image reconstruction, some of them are \textbf{problem-specific methods} that require specialized model architectures or algorithms for the task at hand. 
The theoretical understanding of how these techniques enhance a model's generalization ability has been shown by Wang~\etal~\cite{wang2022generalizing}, and empirical results~\cite{adila2022understanding} also show promising improvements in model performance on both out-of-distribution and in-distribution samples.

The general learning objective of data-level DG can be expressed as:
\begin{equation}
\label{eq-augmentation-all}
\min_{h} \lambda_1\mathbb{E}_{\mathbf{x},y} [ \mathcal{L}(h(\mathbf{x}),y) ]+\lambda_2\mathbb{E}_{\mathbf{x}',y'} [ \mathcal{L}(h(\mathbf{x}'),y') ],
\end{equation}
where $\mathcal{S} = \{(\mathbf{x}_j, y_j)\}^n_{j=1}\sim P_{XY}$ refers to the source domain, $\mathcal{S}' = \{(\mathbf{x}'_j, y'_j)\}^n_{j=1}\sim P_{XY}$ refers to manipulated domain derived from the distribution of the source domain $P_{XY}$ using data-level DG methods, and $\lambda$ is a constant hyperparameter.
The parameter $\lambda_1$ determines the extent to which original data contributes to the learning process, and $\lambda_2$ quantifies the influence of manipulated data on the process. Specifically, when $\lambda_1>0$ and $\lambda_2>0$, data augmentation is employed alongside original data, while if $\lambda_1 = 0$, the learning objective function relies exclusively on manipulated data.
Hence, existing data-level DG can further be refined by choosing the manipulated domain $\mathcal{S}'$, resulting in the methods in the following sections.

\subsubsection{Data Manipulation}\label{sec:data-manipulation}
In data manipulation methods, $\mathcal{S}'$ can be defined as a transformed version of the original dataset $\mathcal{S}$, where each sample has been modified using a specific manipulation function.
This manipulation function, $\mathcal{M}(\cdot)$, can be a closed-form or learnable function that alters the characteristics of the data, making the manipulated data different from the source data.
The specific form of the function $\mathcal{M}(\cdot)$ often depends on the data and the task.
To this end, a manipulated domain $\mathcal{S}'$ that encapsulates data manipulation methods could be defined as:
\begin{equation}
    \mathcal{S}' = \{(\mathcal{M}(\mathbf{x}_j), y_j)\}^n_{j=1}\sim P_{XY}.
\end{equation}

\paragraph{Image Processing Methods\label{par:image-processing}}
Image processing techniques involve closed-form or learnable transformation functions to increase the diversity and quantity of training data.
Examples of some traditional image processing methods include registration, resampling, and filtering, which are specifically designed for the distinct characteristics of the medical image data in question.
Although many traditional image processing methods have empirically shown to improve the model's generalizability~\cite{shah2011evaluating}, they are predominantly employed as pre-processing tools for downstream tasks, rather than as standalone solutions for DG.
Also, with the advancement of deep learning, there has been a gradual shift towards incorporating these techniques directly into deep learning architectures, enabling a more seamless integration of end-to-end learning of domain-invariant features (see Section~\ref{sec:feature-level}).
Readers are referred to~\cite{chlap2021review} for a comprehensive review of the traditional image processing methods.
In the following paragraphs, we explore several deep learning-based image processing methods specifically designed for DG for \MIA tasks.

\parggg{Intensity normalization} methods aim to normalize the raw intensity values or their statistics to reduce the impact of variations in image intensity across different domains.
Several deep learning-based works~\cite{panic2023normalization} have been proposed for intensity normalization technique, typically utilizing an autoencoder-based approach.
For example, inspired by z-score normalization, Yu~\etal~\cite{yu2023san} proposed a U-Net-based~\cite{ronneberger2015u} self-adaptive normalization network (SAN-Net) for the stroke lesion segmentation task.
The U-Net encoder of SAN-Net minimizes the inter-site discrepancy by learning the site-invariant representation with a site classifier and a gradient reversal layer, and the decoder outputs an intensity-normalized image that removes any site-related distribution shifts.
Karani~\etal~\cite{karani2021test} proposed an intensity denoising method for medical image segmentation.
The DAE is trained on intensity-perturbed images to produce denoised outputs, which are then used to train a segmentation CNN.

Other image processing techniques often involve applying a linear or non-linear transformation to the image intensities, such as histogram matching and color normalization.
\parggg{Histogram matching} is a contrast adjustment method that scales pixel values to fit the range of a specified histogram.
Ma~\cite{ma2021histogram} showed that augmenting the source domain with histogram-matched images improves generalization performance for the cardiac image segmentation task.
A subsequent benchmark by Li~\etal~\cite{li2021atrialgeneral} also revealed that histogram matching had the highest performance compared to some commonly used DG methods for atrial segmentation.
Gunasinghe~\etal~\cite{gunasinghe2022domain} proposed a randomized histogram matching method for glaucoma detection that sequentially matches a target image's histogram to multiple randomly selected reference images from the source domain.
This process iteratively adjusts the target image's intensity distribution, promoting a better representation of the source domain.

Global \parggg{color normalization}~\cite{shinohara2014statistical} transfers color statistics by globally altering the image histogram, while local color normalization transfers color statistics of specific regions, preserving intensity information within regions of interest.
These color normalization methods are commonly used in histopathology images, and these methods have improved the generalizability of a neural network~\cite{pontalba2019assessing}.
Kondo~\etal~\cite{kondo2022tackling} employed a color normalization method~\cite{vahadane2016structure} in their architecture for mitosis detection in histopathology images.
This color normalization method decomposes the input image into stain density maps and combines them with the stain color basis of a target image.
Xiong~\etal~\cite{xiong2020improve} introduced the Enhanced Domain Transformation, a color transformation method to align the color space distributions of seen and unseen data for diabetic retinopathy classification.
Pakzad~\etal~\cite{pakzad2022circle} introduced a color transformer utilizing StarGAN~\cite{choi2018stargan} to diversify clinical skin images by altering skin types while retaining original visual characteristics, enhancing dataset diversity and reducing skin type biases in skin disease classification.

{\REV It is worth noting that some \parggg{harmonization} techniques, while primarily aimed at reducing non-biological variability across imaging sites or protocols, can be applied in a DG setting. A notable example is the ComBat harmonization~\cite{johnson2007adjusting}, which uses an empirical Bayes framework to adjust for site effects while preserving biological variability. More recently, deep learning-based harmonization techniques have emerged. Those include autoencoders with heavy regularization or normalization layers~\cite{golkov2016q,koppers2019spherical}, and generative adversarial networks (GANs) capable of scanner-to-scanner translation~\cite{zhao2019harmonization,9354797}. Some of those methods, such as adversarial learning with domain classifiers~\cite{guan2021multi,dinsdale2021deep} and conditional variational autoencoders~\cite{moyer2020scanner}, showed the capability of deriving scanner-/cohort-invariant features for reconstructing harmonized samples across unseen scanners.}

\paragraph{Surrogate Methods\label{par:surrogate}}
Surrogate methods involve using a surrogate representation, such as summary statistics or closed-form mathematical representations, as a substitute for the original input data to improve the generalization performance.

One traditional example is the \parggg{frequency-based DG}\label{par:frequency}, which employs Fourier transformation to separate an image into its amplitude and phase components, typically representing style and content, respectively~\cite{xu2023fourier}.
This is motivated by a well-known property of Fourier transformation that amplitude contains low-level statistics while phase contains high-level semantics~\cite{xu2021fourier}.
The goal of frequency-based DG is to manipulate the low-level statistics of the amplitude component without significantly varying the high-level semantics of the phase component.
These methods are usually well-suited for tasks where high contrast is advantageous, such as fundus imaging~\cite{xu2023fourier} or image segmentation tasks~\cite{zhao2022test}.
For the white matter hyperintensity segmentation task, Zhao~\etal~\cite{zhao2022test} creates amplitude prototypes from source domains and learns a calibrating function that reduces the divergence between source and target amplitudes during inference time.
Inspired by Mix-Up~\cite{zhang2017mixup}, Xu~\etal~\cite{xu2023fourier} introduces perturbation to the amplitude by interpolating the amplitudes of images from different domains for the fundus image segmentation task.
Lie~\etal~\cite{liu2022domain} proposed an alternative frequency-based DG for fundus image restoration, which uses a Gaussian filter to decompose low-frequency and high-frequency components from an image.
Hu~\etal~\cite{hu2022domain} uses Hessian matrices of an image for retinal vessel segmentation, as vector fields better capture the morphological features and suffer less from covariate shift.

\phantomsection\label{par:data-raw} 
Distribution shifts in medical imaging often arise from image reconstruction processes, which transform raw device data into interpretable images. An alternative is to train \parggg{using raw signals}, such as $k$-space data in MRI and sinogram-space data in CT, to circumvent domain-specific variations introduced by reconstruction algorithms and scanner parameters.
Lee~\etal~\cite{lee2019machine} found that a sinogram-space CNN was about 3\% more accurate than an image-space CNN in body part recognition tasks, demonstrating the advantage of using sinogram-space data over CT images. Their findings, along with the potential for radiomics signature analysis on raw data~\cite{gallardo2016normalizing}, underscore the benefits of leveraging raw image data to bypass reconstruction biases.
For example, Zakazov~\etal~\cite{zakazov2022feather} proposed a DG method that operates on $k$-space data for brain segmentation tasks.
The proposed method transfers the contrast and structure-related features by swapping the low-frequency areas (\ie, center) of the target $k$-space data with that of the source $k$-space data.
Zhang~\etal~\cite{zhang2022motion} tackled motion correction in brain MRI by training their model on synthesized motion-corrupted images, which created by introducing motion artifacts into the $k$-space data.

\parggg{Dictionary learning}~\cite{zhao2021survey}, or sparse representation learning, can be considered as a type of surrogate method that seeks to find a sparse representation of input data (\ie, the surrogate) as a linear combination of basic elements, capturing common structures while reducing domain-specific variations~\cite{elad2006image}. Song~\etal~\cite{song2019coupled} applied this to multi-contrast MRI reconstruction by learning dictionaries that highlight structural similarities. Similarly, Liu~\etal~\cite{liu2022single} used dictionary learning for prostate MRI and fundus image segmentation, constructing a shape dictionary with templates to represent diverse segmentation masks efficiently.

\subsubsection{Data Augmentation\label{sec:data-augmentation}}
{\REV
Data augmentation is one of the most prevalent and important forms of DG in \MIA. It refers to techniques that artificially expand and diversify the training dataset by applying various transformations to existing data. The primary goal is to improve the model's ability to generalize across different domains by exposing it to a wider range of data variations during training. Unlike feature-level augmentation (\secref{sec:feat_aug}), which modifies the learned feature representations, data-level augmentation directly alters the input data-space.

The widespread adoption and significance of data augmentation in DG for \MIA stem from its effectiveness, relative simplicity, and broad applicability across different tasks and modalities. It serves several crucial purposes:
\begin{itemize}
    \item Simulating domain shift: By applying transformations that mimic potential variations across different domains (e.g., changes in image intensity, contrast, or noise levels), models can learn to be more robust to these shifts.
    \item Addressing data scarcity: In medical imaging, where large, diverse datasets are often challenging to obtain, augmentation can help mitigate the limitations of small sample sizes.
    \item Enhancing model generalizability: By exposing the model to a broader range of data variations, it can learn more robust and generalizable features.
\end{itemize}

For a comprehensive review of general (non-DG) data augmentation methods for \MIA, readers are referred to the survey by Chlap~\etal~\cite{chlap2021review}. In the following subsections, we focus on augmentation techniques specifically designed or adapted for domain generalization in medical image analysis.}

\paragraph{Randomization-based Augmentation}\label{par:data-randomization}
The idea of random augmentation is to generate novel input data by applying random transformations to the original data.
Some conventional techniques include randomly applying flipping, rotation, scaling, cropping, adding noise, \textit{etc.}, which are used extensively to improve a model's generalization performance by reducing overfitting~\cite{shorten2019survey}. Li~\etal~\cite{li2021random} developed a novel style transfer network that augments a domain by modifying cardiac images with randomly sampled shape and spatial (\ie, slice index) priors to alleviate the modality-level difference for cardiac segmentation.
Liu~\etal~\cite{liu2022domain} proposed a random amplitude mixup method that randomly mixes the amplitudes of different images for DG for fundus image restoration.
{\REV Billot~\etal~\cite{BILLOT2023102789} introduced SynthSeg, a novel approach that leverages domain randomization to train a segmentation network on synthetic brain MRI scans with randomized contrasts and resolutions. This method enables the network to generalize to real scans of varying imaging characteristics without retraining. Similarly, Shen~\etal~\cite{10.1007/978-3-031-16434-7_21} proposed RandStainNA, which unifies stain normalization and augmentation techniques by randomly generating virtual stain templates.}

\paragraph{Adversarial-based Augmentation}\label{par:data-adversarial} Adversarial-based data augmentation methods operate on the principle of creating
adversarial examples that aim to maximize the model’s uncertainty, thereby improving its robustness and generalizability.
In this section, we concentrate on data-level adversarial augmentation, while a discussion on model-level adversarial training can be found in Section \ref{par:model-adversarial}.
Tomar~\etal~\cite{tomar2023tesla} developed a method that combines knowledge distillation with adversarial-based data augmentation for cross-site medical image segmentation tasks.
The process involves the creation of augmented data that is adversarial to the current model, to push the model's feature representations toward the decision boundary.
This is achieved by optimizing and sampling data augmentations that simulate data in the uncertain region of the feature space, thereby improving the model's ability to generalize from the training data to unseen test data.

\paragraph{Generative Models}\label{par:data-generative}
Generative models have been widely used for data augmentation in DG tasks.
These models learn to generate new data that mirrors the training data distribution, thus providing additional examples for the model to learn from.
Scalbert~\etal~\cite{scalbert2022test} designed a new augmentation strategy based on multi-domain image-to-image translation to enhance robustness in unseen target protocols. By adapting the style encoding method~\cite{choi2020stargan} based on generative models, they derive a considerable boost of performances for DG at test time.
Yamashita~\etal~\cite{yamashita2021learning} proposed a style transfer-based augmentation (STRAP) method for a tumor classification task, which applies the style of non-medical images to histopathology images while preserving their semantic content.
The authors argue that the style of these images is specific to their domain and irrelevant to their classification, making STRAP effective in learning domain-agnostic representations.

\subsubsection{Problem-specific Data-level Methods}\label{sec:data-problem-specific}
Problem-specific manipulation methods are tailored to address unique challenges posed by particular types of medical imaging data.

\paragraph{Cross-modal Generative Models}
Cross-modal generative models represent a pioneering paradigm for achieving DG, wherein models are trained to gain knowledge of the data distribution across diverse modalities (\eg, CT, MRI, X-ray, and PET).
These models, often based on GANs, generate synthetic data~\cite{li2021random} or suitable latent representations~\cite{xu2021generative}, which bridge the distributional gap among cross-modalities.
This strategy allows us to provide a model especially capable in medical imaging where data could vary greatly due to patient cohorts, hospital practices, or different imaging modalities. As obvious advantages of such a model, it can be highly valuable when one modality is unavailable for a particular patient or when the model is required to generalize to an unseen domain where a different imaging modality is used.
Readers are {\REV referred} to Xie~\etal~\cite{10.1145/3625227} for a comprehensive review on cross-modal neuroimage synthesis.

Taleb~\etal~\cite{taleb2021multimodal} introduced a self-supervised learning strategy using multimodal jigsaw puzzles for synthesizing cross-modal medical images, where patches from different imaging modalities are assembled to enhance feature extraction across modalities. They further augmented multimodal data volume by generating synthetic images between modalities through a CycleGAN-based translation model.
Xu~\etal~\cite{xu2022adversarial} proposed an adversarial domain synthesizer for single-source cross-modality image segmentation, employing adversarial training coupled with a mutual information regularizer to maintain semantic consistency between original and synthetic domains. 
Su~\etal~\cite{su2023rethinking} introduced the Saliency-balancing Location-scale Augmentation (SLAug) for enhancing cross-modal and cross-sequence medical image segmentation. SLAug modifies image distribution with class-specific adjustments and dynamically tunes location-scale weights via model gradients, effectively mitigating domain shifts in medical imaging.

\paragraph{Stain normalization} Stain normalization and stain separation techniques are primarily used in histopathology, where different tissue components (\eg, nuclei, cytoplasm, extracellular matrix) are separated based on their staining patterns. This process helps remove staining artifacts and enhances the precision of \MIA tasks, such as cell counting and segmentation.
Xu~\etal~\cite{xu2022improved} proposed a stain normalization method for cell detection in histopathology images.
Specifically, the authors address the limitations of stain transformation performed during network training, which may not perfectly represent the stain color of test images. 
Thus, their approach involves mixing stain colors of target and source domain images and generating multiple transformed test images for better stain representation during testing.
Chang~\etal~\cite{chang2021stain} proposed Stain Mix-Up for the cancer detection task.
By decomposing histopathology images into stain color matrices and density maps, the stain mix-up method allows for combining stain colors from different domains.
This approach enhances the color diversity in the training data, improving cancer detection performance.
The stain mix-up technique can effectively address stain color variations and staining artifacts, providing more accurate and reliable results for histopathology image analysis.

%% file: tables/data_level_dg.tex
\begin{table*}[htbp]\scriptsize\setlength{\tabcolsep}{7.pt}
    \caption{Data-level domain generalization methods. Methods categorized by different settings for source and target domains (see \secref{sec:settings}), task, organ, and modality used in experiment.}
    \label{tab:data-level}
    \centering
    \begin{tabular}{C{2.4cm}C{1.5cm}C{0.6cm}C{1.5cm}C{1.8cm}C{1.8cm}C{1.8cm}C{1.8cm}}
    \toprule
     {\bfseries Method} & {\bfseries Specific} & {\bfseries Ref.} & {\bfseries Source} & {\bfseries Target} & {\bfseries Task} & {\bfseries Organ} & {\bfseries Modality}\\
\toprule
    Image Processing & \mrow{1}{Intensity Normalization} & \cite{yu2023san} & Multiple & Site & Segmentation & Brain & MRI\lc{f:atlas}\\\ccmrule
     & & \cite{karani2021test} & Single & Site, Sequence & Segmentation & Brain\lc{f:hcp,f:abide}, Prostate\lc{f:promise,f:nci-isbi}, Cardiac\lc{f:acdc,f:rvsc} & MRI\\
     \cmrule
     & \mrow{1}{Histogram Matching} & \cite{ma2021histogram} & Multiple & Site & Segmentation & Cardiac & MRI\lc{f:mms}\\\ccmrule
     &  & \cite{li2021atrialgeneral} & Multiple & Site & Segmentation & Atrial & MRI \lc{f:cdemris,f:asc}\\\ccmrule
     &  & \cite{gunasinghe2022domain} & Multiple & Site & Segmentation & Retinal & Fundus\lc{f:rim-one-r3,f:refuge} \\
     \cmrule
     & \mrow{1}{Color Normalization} & \cite{kondo2022tackling} & Multiple & Site & Detection & Tissue & Histology\lc{f:midog}\\\ccmrule
     & & \cite{xiong2020improve} & Multiple & Site & Classification & Retinal & Fundus \\\ccmrule
     & & \cite{pakzad2022circle} & Multiple & Site & Classification & Skin & Dermatology\lc{f:fitzpatrick17k} \\\ccmrule
     & & \cite{zhang2022semi} & Single, Multi & Site & Segmentation, Classification & Retinal, Chest & Fundus\lc{f:rim-one-r3,f:drishti-gs,f:refuge}, X-ray\lc{f:mimic-cxr,f:cxr8,f:chexpert,f:padchest} \\
     \midrule
     Surrogate & \mrow{1}{Frequency-based DG} & \cite{zhao2022test} & Multiple & Site & Segmentation & Brain & MRI\lc{f:wmh}\\\ccmrule
     & & \cite{xu2023fourier} & Single, Multiple & Site & Segmentation & Retinal & Fundus\lc{f:refuge}\\\ccmrule
     & & \cite{liu2022domain} & Multiple & Site & Restoration &  Retinal & Fundus\lc{f:kag-cat,f:drive}  \\\ccmrule
     & & \cite{hu2022domain} & Multiple & Site & Segmentation & Retinal & Fundus\lc{f:drive}, OCT\lc{f:octa-500,f:rose} \\\ccmrule
     & & \cite{li2023frequency} & Single & Site & Segmentation & Retinal & Fundus\lc{f:drive,f:kdr,f:iostar,f:les-av} \\    
     \cmrule
     & \mrow{1}{Using Raw Signals} & \cite{zakazov2022feather} & Multiple & Site & Segmentation & Brain & MRI\lc{f:cc359} \\\ccmrule
     & & \cite{zhang2022motion} & Single & Sequence & Segmentation & Brain & MRI \\
     \cmrule
     & \mrow{1}{Dictionary Learning} & \cite{song2019coupled} & Multiple & Sequence & Reconstruction & Brain & MRI\lc{f:brainweb}\\\ccmrule
     & & \cite{liu2022single} & Single & Site & Segmentation & Prostate, Retinal & MRI\lc{f:promise,f:i2cvb,f:nci-isbi}, Fundus\lc{f:drishti-gs,f:rim-one-r3,f:refuge} \\
    \midrule
     Augmentation & \mrow{1}{Randomization-based} & \cite{li2021random} & Single & Sequence & Segmentation & Cardiac & MRI \lc{f:mms} \\\ccmrule
     & & \cite{liu2022domain} & Multiple & Site & Restoration &  Retinal & Fundus\lc{f:kag-cat,f:drive} \\\ccmrule
     & & \cite{zhang2022robust} & Single & Site & Segmentation & Retinal & Fundus\lc{f:drishti-gs,f:rim-one-r3,f:refuge} \\    \ccmrule
     & & \cite{gomathi2022novel} & Single & Site & Segmentation & Retinal & Fundus \\
     \cmrule     
      & Adversarial-based & \cite{tomar2023tesla} & Multiple & Site & Segmentation & Spinal\lc{f:scgm}, Prostate\lc{f:scgm,f:i2cvb,f:nci-isbi}, Colon\lc{f:kather16,f:kather18} & MRI, Histology  \\
     \cmrule
     & \mrow{1}{Generative Model}& \cite{scalbert2022test} & Multiple & Site & Classification & Colon\lc{f:kather16,f:crc-tp,f:kather18}, Breast\lc{f:camelyon17} & Histology \\\ccmrule
     & & \cite{yamashita2021learning} & Single & Site & Classification & Colon & Histology\lc{f:kather16,f:kather18} \\    \ccmrule
     & & \cite{yang2023learning} & Single & Sequence, Site & Detection & Liver & PET, CT \\    
\midrule

     Problem-specific & \mrow{1}{Cross-modal Generative Model} & \cite{taleb2021multimodal} & Multiple & Modality, Sequence & Segmentation & Brain\lc{f:brats}, Prostate, Abdominal\lc{f:chaos} & MRI, CT \\\ccmrule
     
      & & \cite{xu2022adversarial} & Single & Modality, Sequence & Segmantation & Prostate\lc{f:promise,f:i2cvb,f:nci-isbi}, Abdominal\lc{f:sabsct,f:chaos}, Cardiac\lc{f:ms-cmrseg} & MRI, CT\\\ccmrule
     & & \cite{su2023rethinking} & Single & Modality  &  Segmentation & Cardiac\lc{f:ms-cmrseg}, Abdominal\lc{f:sabsct,f:chaos} & MRI, CT \\
    \cmrule
     
      & \mrow{1}{Stain Normalization}& \cite{xu2022improved} & Multiple & Site & Detection & Breast & Histology\lc{f:tupac} \\\ccmrule
     & & \cite{chang2021stain} & Multiple & Site & Classification, Segmentation & Tissue, Breast\lc{f:camelyon17} & Histology \\
     
\bottomrule
\end{tabular}
\end{table*}

%% file: sections/methods/feature.tex
\subsection{Feature-level Generalization}
\label{sec:feature-level}
\input{tables/feature_level_equations}
\input{tables/feature_level_dg}
Feature-level generalization methods aim to utilize the domain-invariant features from the input images to improve the generalization performance of a model.
These methods often involve learning a feature representation shared across multiple domains, either by training a domain-invariant feature extractor or adapting the feature extractor on the fly during inference.
We denote $f$ as a feature mapping function that maps input data to a feature space.
The objective function of domain generalization from Eq.~\ref{eq:overall} can be modified to include a feature extractor $f: \mathcal{X} \to \mathcal{Z}$ and the redefined predictive function $h: \mathcal{Z} \to \mathcal{Y}$:
\begin{equation}\label{eq:feature-level}
    \min_{h, f} \, \mathbb{E}_{(\mathbf{x},y)} [ \mathcal{L}(h(f(\mathbf{x})),y) ].
\end{equation}
Refer to Table~\ref{tab:feature-level-methods} for a summary of feature-level methods.
In the following paragraphs, we explore feature-level domain generalization techniques.

\subsubsection{Feature Alignment\label{sec:feature-alignment}}
Feature alignment aims to align or standardize the feature distributions across different domains.
These strategies aim to produce domain-invariant features through statistical and structural adjustments, enhancing generalization across varied domains by minimizing distributional discrepancies and aligning feature distributions to a common representation.

\paragraph{Feature Normalization}
Feature normalization methods aim to statistically center, scale, decorrelate, standardize, or whiten feature distributions across domains and enhance the model's ability to generalize~\cite{huang2023normalization}.
By transforming all features to the same statistical distribution, normalization prevents features with larger numerical values from dominating those with smaller ones during training, ensuring a more balanced and accurate model.
These methods generally stem from the traditional scaling methods, such as z-score and unit vector normalization, as well as some traditional machine learning methods, such as batch and instance normalization.
These methods can be formulated as the following generalized equation for feature normalization:
\begin{equation}\label{eq:feature-normalization}
    \min_{h, f} \, \mathbb{E}_{(\mathbf{x},y)} \left[ \mathcal{L}\left(h\left(\frac{\mathbf{z}-\boldsymbol{\mu}}{\sqrt{\boldsymbol{\sigma}^2+\epsilon}}\right),y\right) \right],
\end{equation}
where $\mathbf{z}=f(\mathbf{x})$ is the feature embedding, $\boldsymbol{\mu}$ and $\boldsymbol{\sigma}^2$ are the statistics of the feature embedding $\mathbf{z}$ (usually the mean and variance), and $\epsilon$ is a constant for numerical stability.

Zhou~\etal~\cite{zhou2022generalizable} proposed a per-domain batch normalization method for medical image segmentation.
When testing the model on the target domain, the model compares the distribution information of the target domain with the stored distribution information (mean and variance) from each domain.
Then, the model selects the most suitable domain distribution statistics to normalize the activated features from the target domain.
Liu~\etal~\cite{liu2023ss} introduced spectral-spatial normalization (SS-Norm) for retinal vessel segmentation, merging frequency and spatial normalization to isolate domain-invariant features. The approach uses discrete Fourier transformation for frequency normalization and a convolutional network for spatial normalization, improving the representation of spatial details in activation maps.

\paragraph{Dissimilarity-based Alignment}
Dissimilarity-based alignment methods attempt to reduce the difference between the feature distributions of different domains by minimizing a dissimilarity measure.
This aligns the distributions to a common representation, which helps mitigate the domain shift problem.
The goal of dissimilarity-based alignment is to find $f$ to minimize the distribution shift among domains in the feature space.
For instance, given the $i$-th and $j$-th source domains with input samples $\mathbf{x}^i$ and $\mathbf{x}^j$, we may want to minimize the difference between the distributions of their mapped features: $\mathcal{D}(f(\mathbf{x}^i), f(\mathbf{x}^j))$, where $\mathcal{D}(\cdot, \cdot)$ measures the dissimilarity between two distributions, \ie,
\begin{equation}
\min_{f} \mathcal{D}(f(\mathbf{x}^i), f(\mathbf{x}^j)), \quad 1 \le i \ne j \le M.
\end{equation}
Numerous statistical metrics exist to measure the dissimilarity between distributions, including $\ell_2$ distance, $f$-divergences, and the Wasserstein distance.

Stacke~\etal\cite{stacke2020measuring} empirically evaluated different dissimilarity metrics for tumor classification in cross-site histopathology images.
Among various metrics, Wasserstein-based metrics have been shown to better capture the domain shift in cross-site histopathology images.
Lyu~\etal~\cite{lyu2022aadg} applied a Wasserstein-based metric, specifically the Sinkhorn distance, to measure divergence between augmented domains created through varied image transformations for retinal image segmentation.
This approach facilitated the evaluation of domain shift through the divergence of novel distributions induced by different augmentation sub-policies.
Similarly, Li~\etal~\cite{li2020domain} developed Linear-Dependency Domain Generalization (LDDG) to improve generalization for lesion classification and spinal cord segmentation by aligning latent feature distributions across multiple source domains using Kullback-Leibler (KL) divergence and linear dependency modeling. This approach seeks to reduce empirical risk on unseen target domains, aiming for a theoretical performance upper bound.

\subsubsection{Disentanglement Methods}\label{sec:feature-disentanglement}
Disentanglement methods aim to decompose an input sample into a feature vector that reveals various factors of variation where each dimension or subset of dimensions carries information linked to a specific factor.
The primary goal of these methods is to create a clear boundary between domain-specific and task-specific features.
This distinction is crucial in capturing the universal patterns related to the task.
Given this goal, the disentanglement process seeks to isolate task-relevant features from those features intrinsic to the domain, \ie, $\mathbf{z} = [\mathbf{z}_{\text{task}}, \mathbf{z}_{\text{domain}}]$, respectively.
The goal is to create a model that emphasizes $\mathbf{z}_{\text{task}}$ while effectively ignoring $\mathbf{z}_{\text{domain}}$, thus ensuring that the model's focus is primarily on the features that contribute to the task at hand and less on those that are domain-specific features.
To this end, we further refine disentanglement methods into implicit and explicit methods.

\paragraph{Implicit Feature Disentanglement}
Implicit feature disentanglement strategies learn to decompose factors of variations by utilizing, for example, the statistical properties of the data and indirect incentives to encourage disentanglement.
Such approaches provide scalable and flexible techniques for learning disentangled representations.
Typical examples of these methods include information-theoretic methods, contrastive learning, and variational inference.

\parggg{Information theoretic disentanglement} methods often focus on using mutual information to separate and understand the different factors of variations in data.
Mutual information, denoted by $I(X;Y)$, measures the information obtained from a random variable $X$ by observing another variable $Y$.
The goal of information-theoretic disentanglement is to minimize the mutual information between the task and domain representations, \ie, 
\begin{equation}
    \min_f I({\mathbf{z}}_{\text{task}}; {\mathbf{z}}_{\text{domain}}),
\end{equation}
where $f\left(\mathbf{x}\right)=\left[{\mathbf{z}}_{\text{task}}, {\mathbf{z}}_{\text{domain}}\right]$ is a feature mapping function that disentangles the input image into ${\mathbf{z}}_{\text{task}}$ and ${\mathbf{z}}_{\text{domain}}$.
This minimization process plays a vital role in ensuring that the task-related and domain-specific feature sets are \emph{independently} informative.
This disentanglement approach seeks to construct a learning model capable of robustly interpreting and classifying data across a spectrum of domains, making it adaptable to a wide range of task-specific challenges in diverse applications.

Specifically, Meng~\etal~\cite{meng2020mutual} proposed MIDNet, an MI-based model specifically designed for fetal ultrasound classification tasks. MIDNet's primary objective is to distinguish domain-invariant features from domain-specific ones by minimizing the mutual information between these feature sets.
To achieve this, they employ the Mutual Information Neural Estimation (MINE)~\cite{belghazi2018mutual} approach to approximate the lower bound of the mutual information. This facilitates the extraction of generalizable features and enables knowledge transfer across unseen categorical features in target domains.
Similarly, Bi~\etal~\cite{bi2023mi} proposed MI-SegNet for ultrasound image segmentation.
MI-SegNet employs two encoders that separately extract anatomical and domain features from images, and MINE approximation is used to minimize the mutual information between these features.
Rather than minimizing the mutual information between domains, Chen~\etal~\cite{chen2021d} and Xu~\etal~\cite{xu2022adversarial} proposed to maximize the mutual information for maintaining the consistency between the source domain and augmented samples.

\parggg{Contrastive Disentanglement}\label{par:feature-contrastive} aims to make representations of similar instances more alike (low contrast) and those of different instances more dissimilar (high contrast).
{\REV In the context of domain generalization, this approach can be used to learn domain-invariant features by encouraging similarity between samples that share the same task-relevant characteristics, regardless of their domain, while separating samples with different characteristics.}
A typical contrastive learning loss function~\cite{oord2018representation} is defined as:
\begin{equation}\label{eq:contrastive}
\mathcal{L}(\mathbf{x}_i, \mathbf{x}_j)=-\log \frac{\exp \left(\operatorname{sim}\left(f(\mathbf{x}_i), f(\mathbf{x}_j)\right) / \tau\right)}{\sum_{k\in a(i)} \exp \left(\operatorname{sim}\left(f(\mathbf{x}_i), f(\mathbf{x}_k)\right) / \tau\right)},
\end{equation}
where $\operatorname{sim}(\cdot,\cdot)$ is a function for cosine similarity, $\mathbf{x}_i$ and $\mathbf{x}_j$ are positive pairs, $k\in a(i)$ are the indexes of selected negative samples, and $\mathbf{x}_i$ and $\mathbf{x}_k$ are negative pairs.
{\REV In the context of domain generalization for medical image analysis, positive pairs could be defined as images showing the same pathology (\eg, two images of malignant tumors) regardless of whether they come from different hospitals or were acquired using different imaging protocols. Negative pairs would be images showing different pathologies (\eg, an image of a malignant tumor paired with an image of healthy tissue), again regardless of their source domain. This pairing strategy encourages the model to learn features that distinguish between pathologies while being invariant to domain-specific characteristics like image acquisition settings or hospital-specific protocols.}

Li~\etal~\cite{li2021domain} proposed a novel approach that couples multi-style and multi-view contrastive learning to enhance the generalization capability for mammography lesion detection.
Specifically, positive pairs for multi-style contrastive learning were synthesized using a GAN, and different views of the breast (\ie, craniocaudal and mediolateral oblique) were used as multi-view contrastive learning.
In a similar approach, Gu~\etal~\cite{gu2022contrastive} proposed Contrastive Domain Disentanglement and Style Augmentation (CDDSA) for image segmentation in the fundus and MR images.
The unique feature of CDDSA is its implementation of a style contrastive loss function, which ensures that style representations from the same domain bear similarity while those from different domains diverge significantly.

\parggg{Variational disentanglement\label{para:variational}} is a method that utilizes variational autoencoders (VAEs) to learn a disentangled representation.
The typical approach for this method involves encoding input data $\mathbf{x}$ into a latent variable $\mathbf{z}$ using an encoding function $q_\phi(\mathbf{z}|\mathbf{x})$.
The decoder, $p_\theta(\mathbf{x}|\mathbf{z})$, then reconstructs the original data from the latent representation $\mathbf{z}$.
The objective function of VAEs, or the evidence lower bound (ELBO), can be expressed as:
\begin{align}
    \mathcal{L}(\theta, \phi; \mathbf{x}) &= \mathbb{E}_{q_\phi(\mathbf{z}|\mathbf{x})}[\log p_\theta(\mathbf{x}|\mathbf{z})] - \text{KL}(q_\phi(\mathbf{z}|\mathbf{x})||p(\mathbf{z}))\nonumber\\
    &=\mathcal{L}_{rec}+\lambda\mathcal{L}_{reg},
    \label{eq:ELBO}
\end{align}
where $\text{KL}(q_\phi(\mathbf{z}|\mathbf{x})||p(\mathbf{z}))$ is the K divergence between the approximate posterior $q_\phi(\mathbf{z}|\mathbf{x})$ and the prior $p(\mathbf{z})$, which is often chosen to be a normal distribution.
ELBO can also be interpreted as minimizing the reconstruction error $\mathcal{L}_{rec}$, \ie, the posterior $p_\theta(\mathbf{x}|\mathbf{z})$, and regularizing the approximate posterior $\mathcal{L}_{reg}$, \ie, the KL term. 
The key idea behind variational disentanglement involves structuring a latent space so that distinct dimensions capture domain-specific and domain-invariant factors.
This is typically achieved by introducing tailored constraints or regularization mechanisms during training~\cite{9352543}.
For example, regularization or constraints can be incorporated into the ELBO to specifically encourage the separation of domain-specific and domain-invariant factors in the latent space.

Ilse~\etal~\cite{ilse2020diva} proposed the Domain Invariant Variational Autoencoder (DIVA) for malaria cell image classification~\cite{bandi2018detection}.
DIVA is an extension to the VAE framework that can partition a latent space into three independent latent subspaces for domain label $\mathbf{z}_d$, class label $\mathbf{z}_y$, and residual variations $\mathbf{z}_x$, which captures any residual variations left in data $x$.
This partitioning aims to encourage the model to disentangle these sources of variation.
Specifically, DIVA employs three separate encoders that serve as variational posteriors over the three latent variables.
In addition to the ELBO term, DIVA formulates classifier-based auxiliary objectives to further encourage the separation of domain-specific and class-specific information into their respective latent variables:
\begin{align}
&\mathcal{L}(\theta, \phi; \mathbf{x})= \mathbb{E}_{q_{\phi_d}(\mathbf{z}_d | \mathbf{x})q_{\phi_x}(\mathbf{z}_x|\mathbf{x}),q_{\phi_y}(\mathbf{z}_y|\mathbf{x})} \left[ \log p_\theta(\mathbf{x}|\mathbf{z}_d, \mathbf{z}_x, \mathbf{z}_y) \right]\nonumber \\ & \qquad -\beta KL\left(q_{\phi_d}(\mathbf{z}_d|\mathbf{x})||p_{\theta_d}(\mathbf{z}_d|d)\right)
- \beta KL\left(q_{\phi_x}(\mathbf{z}_x|\mathbf{x})||p(\mathbf{z}_x)\right) \nonumber \\
& \qquad -\beta KL\left(q_{\phi_y}(\mathbf{z}_y|\mathbf{x})||p_{\theta_y}(\mathbf{z}_y|y)\right).
\end{align}
Wang~\etal~\cite{wang2021variational} introduced the Variational Disentanglement Network (VDN) for breast cancer metastasis classification, which separates domain-invariant and domain-specific features by maximizing information gain and posterior probability. Through adversarial training between a task-specific encoder and a feature discriminator, VDN aligns latent features with a predefined prior and employs a generator network for high-quality reconstruction and effective feature disentanglement, enhancing domain generalization.
Wang~\etal~\cite{wang2022domain,wang2023learning} propose a variational causal model for the breast cancer classification task.
Specifically, they propose a structural causal model that can decompose the latent factors of medical images into domain-agnostic causal features and domain-aware features.
These features are factored into a reformulated ELBO term of VAE, and optimizing the modified ELBO provably disentangles the domain-agnostic causal features from domain-aware features.

\paragraph{Explicit Feature Disentanglement} There is an explicit mechanism separating task-relevant features from domain-specific features in disentanglement. 
These methods often involve supervision or hard constraints in the model.
Supervision could take the form of domain labels or auxiliary attributes indicating the values of factor of variations for each data instance.
Some methods use constraints or regularization terms in the objective function to encourage the model to separate specific factors of variation in the representations.
The loss for these types of methods can be in the form of: 
\begin{equation}
    \min_{h,f} \mathbb{E}_{(\mathbf{x},c) \in \mathcal{S}} [ \mathcal{L}(h(f(\mathbf{x})),c) ] + \lambda \, \mathcal{L}_{reg}, 
\end{equation}
where $c$ is an auxiliary attribute or a domain label, $\mathcal{L}_{reg}$ is a regularization term that encourages separation between the task-relevant and domain-specific features, and $\lambda$ is a hyperparameter controlling the strength of this regularization.
The first term in this loss refers to model supervision with an auxiliary attribute or a domain label, while the second term encourages the model to keep the task-relevant and domain-specific features separate.

\parggg{Conditional representation learning} refers to learning a representation of the input data influenced by a certain conditioning variable.
This variable can be any additional information, such as domain labels or induced priors.
Conditional representation learning aims to create representations that are sensitive to the specific aspects of the data relevant to the condition, and invariant or insensitive to other aspects.
This can improve performance on tasks where certain aspects of the data are more relevant than others, or where the relevance of different aspects varies under different conditions.

Liu~\etal~\cite{liu2021recursively,liu2022ordinal} proposed the Recursively Conditional Gaussian (RCG) prior for diabetic retinopathy and congenital heart disease diagnosis task.
Their proposed method utilizes the ordinal structure of the class labels to construct an appropriate RCG before the class-related latent space.
This RCG prior enforces a poset constraint that aligns the extracted latent vectors with the ordinal class labels.
By conditioning the latent space on the ordinal labels, the RCG prior aims to learn a representation sensitive to the relevant aspects of the data for the specific diagnosis task, while invariant to other aspects.
Wang~\etal~\cite{wang2020dofe} proposed Domain-oriented Feature Embedding (DoFE) for fundus image segmentation, which incorporates a domain knowledge pool to learn the domain prior information extracted from the multi-source domains.
This domain prior knowledge is then dynamically enriched with the image features to make the semantic features more discriminative.

\parggg{Feature regularization} methods focus on incorporating regularization terms into the learning objective to guide the model toward extracting meaningful and generalizable features.
These methods often utilize penalties that discourage the model from relying too heavily on individual features or encourage the model to maintain certain structures or properties in the learned representations.
Additionally, regularization can be used to encourage the model to learn representations invariant to certain transformations of the data, such as translations or rotations.
These kinds of regularization can make the learned features more robust to data variations that are irrelevant to the task at hand.
For example, this might be done by promoting sparse representations (\eg, dropout~\cite{srivastava2014dropout}, $\ell_1$, $\ell_2$ regularization), where the model is encouraged to use as few features as possible to achieve its task, or by promoting orthogonality, where the model is encouraged to learn features that are independent of each other.

Islam and Glocker~\cite{islam2022frequency} proposed Frequency Dropout (FD) for cardiac image segmentation task.
FD uses a random feature map filtering approach that works as a form of feature-level regularization during training.
In this method, random filters (\eg, Gaussian smoothing, Laplacian of Gaussian, and Gabor filtering) are applied to the feature maps to prevent the neural network from learning frequency-specific image features.
Nguyen~\etal~\cite{nguyen2023adversarially} introduced the Adversarially-Regularized Mixed Effects Deep learning (ARMED) for Alzheimer's disease diagnosis and cell image classification tasks.
ARMED incorporates a regularization mechanism that enforces the model to learn features invariant to specific clusters in the data.
This is achieved by introducing an adversarial classifier that attempts to predict the cluster membership based on the learned features, while the main model is penalized for enabling this prediction.
Wang~\etal~\cite{wang2021embracing} proposed Knowledge Distillation for Domain Generalization (KDDG) for MRI gray matter segmentation task.
KDDG applies a form of feature-level regularization that encourages the student model's predictions to align with the teacher's predictions, thus improving the student model's robustness and generalization capability.

\subsubsection{Other Representation Learning Methods}\label{sec:feature-others}
\paragraph{Feature Augmentation\label{sec:feat_aug}}
Feature augmentation is a technique used to improve machine learning models' generalization capability by transforming the feature space, rather than the input space. Unlike traditional data augmentation, which directly manipulates raw data, feature augmentation operates on the derived features extracted from the raw data.
While data augmentation creates a more comprehensive and diverse source domain by introducing variations at the data level, it is limited by the extent and variety of feasible and meaningful transformations on the raw data.
On the other hand, by working directly in the feature space, feature augmentation allows for a richer set of transformations.
Feature augmentation can also incorporate domain knowledge more effectively, as transformations can be designed to specifically target and vary important features.

Chen~\etal~\cite{chen2022maxstyle} proposed a novel feature augmentation framework, MaxStyle, for cardiac MRI segmentation. MaxStyle introduces adversarial noise into the feature styles and conducts a worst-case style composition search through adversarial training.
This approach broadens the range of augmented styles and makes the model more robust by exposing it to harder cases.
Zhou and Konukoglu~\cite{zhou2023fedfa} proposed a Federated Feature Augmentation (FedFA) for cross-site prostate MRI segmentation.
FedFA augments the features by estimating a vicinity distribution at each layer of the neural network during training, thus enhancing the data representation at each client.
It manipulates the channel-wise statistics of the features, such as the mean and standard deviation, which often carry significant domain-specific information.

\paragraph{Kernel-based Learning}
Kernel-based methods are a classic and effective approach within feature-level domain generalization.
{\REV These methods improve generalization by mapping the original input features into a higher dimensional space. This mapping offers several advantages for domain generalization. 
In the higher-dimensional space, kernel methods can potentially reveal domain-invariant structures that are not apparent in the original feature space.
Kernel methods can also model non-linear relationships in the data, which is particularly useful for capturing intricate patterns in medical images that may be consistent across domains.}
There are various kernel-based methods for feature-level domain generalization, including Support Vector Machine (SVM) variants, Maximum Mean Discrepancy (MMD), and Transfer Component Analysis (TCA). Kernel trick enables these methods to operate in high-dimensional spaces without explicitly calculating the coordinates of the data in that space, but by simply computing the dot products between the images of all data pairs in the feature space. This makes the calculations more tractable and efficient.
These kernel-based methods can benefit medical image analysis as they can handle high-dimensional data and discover complex patterns. They also offer an excellent way to incorporate domain knowledge, such as spatial relationships in images, by defining appropriate kernels.

Wang~\etal~\cite{wang2022embracing} proposed a kernel-based binary classifier for cross-site brain disease diagnosis tasks.
In the kernel setting, we can reformulate the regularization term as:
\begin{equation}
    \mathcal{L}_{reg}=\frac{||f(\mathbf{x})||^2}{2}=\frac{1}{2}\sum_{i} k(\mathbf{x}^i,\mathbf{x}),
\end{equation}
where the norm is the Reproducing kernel Hilbert space (RKHS) norm, and $k(\cdot,\cdot)$ is the kernel function that measures the similarity between two variables.
The RKHS norm captures the classifier's complexity or ``smoothness'' within the chosen kernel space.
The authors use this kernel-based classifier to measure the disharmony and utilize it to improve the generalizability of the given model.
Ayodele~\etal~\cite{ayodele2020supervised} proposed a multi-TCA approach for epileptic seizure detection using an EEG dataset.
In contrast to utilizing the disharmony~\cite{wang2022embracing}, the authors use the RKHS norm to measure the shared subspace between source domains.
Then, they utilize various dimension reduction techniques to extract a generalized feature vector for a recurrent neural network.

%% file: tables/feature_level_equations.tex
\begin{table}[htbp]\scriptsize
    \centering
    \caption{Summary of feature-level domain generalization methods. }
    \label{tab:feature-level-methods}    \resizebox{1\linewidth}{!}{
    \begin{tabular}{c c c}
    \toprule
    \textbf{Method} & \textbf{Formulation} \\
    \toprule
    Normalization & $\hat{\mathbf{z}}=\frac{f(\mathbf{x})-\boldsymbol{\mu}}{\sqrt{\boldsymbol{\sigma}^2+\epsilon}}$ \\\cmruleeq
    Dissimilarity-based & $\min_{f} D(f(\mathbf{x}^i), f(\mathbf{x}^j)), \quad \forall 1 \le i \ne j \le M$ \\\cmruleeq
    Information theoretic & $\min_f I({\mathbf{z}}_{\text{task}}; {\mathbf{z}}_{\text{domain}})$ \\\cmruleeq
    Contrastive & $\mathcal{L}(\mathbf{x}_i, \mathbf{x}_j)=-\log \frac{\exp \left(\operatorname{sim}\left(f(\mathbf{x}_i), f(\mathbf{x}_j)\right) / \tau\right)}{\sum_{k\in a(i)} \exp \left(\operatorname{sim}\left(f(\mathbf{x}_i), f(\mathbf{x}_k)\right) / \tau\right)}$ \\\cmruleeq
    Variational & $\mathcal{L}(\theta, \phi; \mathbf{x}) = \mathbb{E}{q\phi(\mathbf{z}|\mathbf{x})}[\log p_\theta(\mathbf{x}|\mathbf{z})] - \text{KL}(q_\phi(\mathbf{z}|\mathbf{x})||p(\mathbf{z}))$ \\\cmruleeq
    Explicit & $\min_{h,f} \mathbb{E}_{(\mathbf{x},c) \in \mathcal{S}} [ \ell(h(f(\mathbf{x})),c) ] + \lambda \, \ell_{reg}$\\
    \bottomrule
    \end{tabular}
    }
\end{table}

%% file: tables/feature_level_dg.tex
\begin{table*}[htbp]\scriptsize\setlength{\tabcolsep}{7.pt}
    \caption{Feature-level domain generalization methods. Methods categorized by different settings for source and target domains (see \secref{sec:settings}), task, organ, and modality used in experiment.}
    \label{tab:feature-level}
    \centering
    \begin{tabular}{C{2.2cm}C{1.8cm}C{0.6cm}C{1.5cm}C{1.8cm}C{1.8cm}C{1.8cm}C{1.8cm}}
    \toprule
     {\bfseries Method} & {\bfseries Specific} & {\bfseries Ref.} & {\bfseries Source} & {\bfseries Target} & {\bfseries Task} & {\bfseries Organ} & {\bfseries Modality}\\
\toprule
\mrow{1}{Feature Alignment} & \mrow{1}{Feature Normalization} & \cite{zhou2022generalizable} & Single & Sequence, Modality & Segmentation & Brain\lc{f:brats}, Cardiac\lc{f:mm-whs}, Abdominal\lc{f:sabsct,f:chaos} & MRI, CT\\\ccmrule
     &  & \cite{liu2023ss} & Single & Site & Segmentation & Retinal & Fundus\lc{f:chase,f:stare,f:iostar,f:drhagis,f:aria,f:drive}\\
     \cmrule
          & \mrow{1}{Dissimilarity-based} & \cite{stacke2020measuring} & Multiple & Site & Classification & Colon\lc{f:aida-lnco}, Breast\lc{f:camelyon17} & Histology\\\ccmrule
     & & \cite{lyu2022aadg} & Multiple & Modality & Segmentation & Retinal & Fundus\lc{f:kdr,f:e-ophtha,f:drishti-gs,f:rim-one-r3,f:refuge,f:idrid,f:chase,f:stare,f:hrf,f:rose}, OCT\lc{f:octa-500,f:rose}\\\ccmrule
     & & \cite{li2020domain} & Multiple & Sequence, Site & Classification, Segmentation & Skin\lc{f:isic,f:ph2,f:derm7pt,f:dermofit,f:ham10000}, Spinal\lc{f:scgm} & Dermatology, MRI \\
\midrule

    \mrow{1}{Implicit Disentanglement} & \mrow{1}{Mutual Information}& \cite{meng2020learning} & Multiple & Site & Classification & Abdominal, Brain, Femur, Lips & Fetal Ultrasound \\\ccmrule
     &  & \cite{meng2020mutual} & Multiple & Site & Classification & Abdominal, Brain, Femur, Lips & Fetal Ultrasound\\\ccmrule
      & & \cite{bi2023mi} & Single & Site & Segmentation & Carotid & Ultrasound\lc{f:splab}\\\ccmrule
     &  & \cite{chen2021d} & Single, Multiple & Sequence & Classification & Blood Cell & Histology\lc{f:bccd,f:bcisc,f:lisc} \\\ccmrule
     & & \cite{xu2022adversarial} & Single & Modality, Sequence & Segmantation & Prostate\lc{f:promise,f:i2cvb,f:nci-isbi}, Abdominal\lc{f:sabsct,f:chaos}, Cardiac\lc{f:ms-cmrseg} & MRI, CT\\
     \cmrule
     & \mrow{1}{Contrastive} & \cite{li2021domain} & Multiple & Site & Detection & Breast & X-ray\lc{f:inbreast}\\\ccmrule
     & & \cite{gu2022cddsa} & Multiple & Site & Segmentation & Retinal & Fundus\lc{f:drishti-gs,f:rim-one-r3,f:refuge} \\
     \cmrule
     & \mrow{1}{Variational} & \cite{ilse2020diva} & Multiple & Site & Classification & Blood Cell & Histology\lc{f:malariascreener} \\\ccmrule
     & & \cite{wang2021variational} & Multiple & Site & Classification & Breast & Histology\lc{f:camelyon17} \\\ccmrule
    & & \cite{wang2022domain} & Multiple & Site & Classification & Breast & X-ray\lc{f:ddsm} \\\ccmrule
     & & \cite{wang2023learning} & Multiple & Site & Classification & Breast & X-ray\lc{f:ddsm}\\
     \midrule

    \mrow{1}{Explicit Disentanglement} &  \mrow{1}{Conditional \mbox{Representation} Learning} & \cite{liu2021recursively} & Multiple & Site & Classification & Retinal & Fundus\lc{f:kdr,f:idrid} \\\ccmrule
    & & \cite{liu2022ordinal} & Multiple & Site & Classification & Retinal & Fundus\lc{f:kdr,f:idrid} \\\ccmrule
    & & \cite{wang2020dofe} & Multiple & Site & Segmentation & Retinal & Fundus\lc{f:drishti-gs,f:rim-one-r3,f:refuge,f:chase,f:stare,f:hrf,f:drive} \\
     \cmrule
    & \mrow{1}{Feature Reguarlization}& \cite{islam2022frequency} & Multiple & Site & Segmentation & Cardiac & MRI\lc{f:mms} \\\ccmrule
    & & \cite{nguyen2023adversarially} & Multiple & Site & Compression, classification & Brain\lc{f:adni}, Skin\lc{f:openlch} & MRI, Histology \\\ccmrule
    & & \cite{wang2021embracing} & Single & Site & Segmentation & Spinal & MRI\lc{f:scgm} \\
     \midrule
    Others & \mrow{1}{Feature Augmentation} & \cite{chen2022maxstyle} & Single, Multiple & Site & Segmentation & Cardiac\lc{f:mms,f:acdc,f:ms-cmrseg}, Prostate\lc{f:promise,f:i2cvb,f:nci-isbi} & MRI \\\ccmrule
    & & \cite{zhou2023fedfa} & Multiple & Site & Segmentation & Prostate & MRI\lc{f:promise,f:i2cvb,f:nci-isbi} \\\ccmrule
    & & \cite{wen2024denoising} & Single & Site & & Liver & CT\lc{f:chaos,f:lits} \\
     \cmrule
    & \mrow{1}{Kernel-based} & \cite{wang2022embracing} & Single & Site & Classification & Brain & MRI\lc{f:istaging} \\\ccmrule
    & & \cite{ayodele2020supervised} & Multiple & Site & Detection & Brain & EEG\lc{f:tusz,f:chb-mit} \\
     
\bottomrule
\end{tabular}
\end{table*}

%% file: sections/methods/model.tex
\subsection{Model-level Generalization}\label{sec:model-level}
{\REV Model-level generalization focuses on enhancing the intrinsic ability of machine learning models to generalize across domains by modifying core aspects of the model itself, including the learning process, model architecture, and optimization techniques.}
Specifically, such strategies encompass several categories of methods: a) \emph{Learning strategy}, which focuses on adequately reflecting the target-suitable knowledge or leveraging distinct representations gained from a variety of sub-tasks; b) \emph{Model framework}, which exploits modifications to the network architecture or the incorporation of adaptive auxiliary components to more efficiently address the domain shift; and lastly c) \emph{Other model-based DG}, which involve various optimization and adaptation techniques.

\input{tables/model_level_dg}
\subsubsection{Learning Strategy\label{sec:model-learning}}
Methods in this category concentrate on harnessing the general learning strategy to enhance the model's generalizability, which mainly involves various techniques as a) \emph{Meta-learning}, wherein the model learns how to rapidly adapt to new tasks, thereby improving its flexibility and generalization capacity; b) \emph{Self-supervised learning}, which is an unsupervised manner that can leverage large amounts of unlabeled data by creating pretext tasks; c) \emph{Adversarial learning}, which strives to minimize the divergence between different domains to enhance the model's transferability.

\paragraph{Meta-learning}
Meta-learning techniques are closely relevant in medical imaging due to the prevalent scarcity of annotated data coupled with the need to rapidly adapt to unseen data domains. Specifically, a model employing such strategies aims to learn an optimal initialization or update rule that can be quickly fine-tuned to perform well in unseen data domains. By virtue of these advantages, it is possible to improve the model's flexibility and the efficiency of its generalization capabilities.
To simulate domain shift, meta-learning methods divide the source domains into meta-training and meta-test sets.
Meta-learning can be formulated as follows: 
{\REV
\begin{align}
\phi^{*} &= \arg\min_{\phi} \mathcal{L}_{meta}(\phi; \mathcal{S}_{mtrain}),  \nonumber\\
\theta^{*} &= \arg\min_{\theta} \mathcal{L}_{task}(\theta; \mathcal{S}_{mtest}, \phi^{}),
\end{align}
where $\phi^{*}$ denotes the meta-learned parameters optimized on the meta-training set $\mathcal{S}_{mtrain}$, which are then used to initialize the task-specific parameters $\theta^{*}$ optimized on the meta-test set $\mathcal{S}_{mtest}$. $\mathcal{L}_{meta}$ and $\mathcal{L}_{task}$ are the meta-learning and task-specific loss functions, respectively. This formulation is inspired by the Model-Agnostic Meta-Learning (MAML) algorithm~\cite{10.5555/3305381.3305498}, where the meta-objective is to find an initialization that allows for quick adaptation to new tasks.
}

Khandelwal and Yushkevich~\etal~\cite{khandelwal2020domain} extended the Meta-learning for Domain Generalization~\cite{li2018learning} for the CT vertebrae segmentation task (MLDG-Seg).
The key idea behind MLDG-Seg is to simulate the domain shift during the training process by artificially creating a meta-test set $\mathcal{S}_{mtest}$ from multiple source domains and then training the model in a way that optimizes its performance across these varied domains or tasks. 
Dou~\etal~\cite{dou2019domain} proposed Model-agnostic learning of Semantic Features (MASF) for the cross-site brain MRI segmentation task.
MASF employs a meta-learning algorithm that enhances generalization to unseen domains by globally aligning class relationships and locally clustering class-specific features, optimizing semantic feature representations.
This approach updates model parameters for improved accuracy in source domains during meta-training, and enforces semantically relevant learning through global and local mechanisms during meta-testing.
Liu~\etal~\cite{liu2020shape} proposed a Shape-aware Meta-learning (SAML) approach for the prostate MRI segmentation task.
SAML introduces two loss functions specifically designed to improve the compactness and smoothness of segmentation in the presence of domain shift.
The compactness loss function encourages segmentations to preserve the complete shape of the prostate, while the smoothness loss function enhances boundary delineation by promoting intra-class cohesion and inter-class separation between contour-relevant and background-relevant embeddings across different domains. 
Lie~\etal~\cite{liu2021semi} proposed a semi-supervised meta-learning approach for domain generalization in medical image segmentation tasks.
Specifically, they split their training dataset into meta-train and meta-test sets, including labeled and unlabeled data, enabling their model to generalize to unseen domains.
Hu~\etal~\cite{hu2023map} proposed Meta-Learning on Anatomy-Consistent Pseudo-Modalities (MAP) for the retinal vessel segmentation tasks.
MAP employs a mixup technique with episodic training on synthesized pseudo-modalities to emphasize structural vessel features, achieving improved generalization across different imaging domains.

\paragraph{Self-supervised Learning}\label{par:model-ssl}
Self-supervised learning (SSL) is a novel learning paradigm where the model is trained to figure out a \textit{pretext} task that learns general but useful feature representations from unlabeled large-scale data. Specifically, the principal idea behind SSL is to design a proxy where the answers can be deduced by a portion of the input data, enabling the model to learn representations under its own supervision.
Thanks to such an advantage, creating the pretext task can alleviate the chronic issues induced by a scarcity of annotated data, especially in medical imaging. Further fine-tuning the downstream task via these universally useful features improves the generalization capability, allowing the model to adequately escape overfitting for domain-specific biases. 
A typical example of SSL is the contrastive learning paradigm introduced in Eq.~\ref{eq:contrastive}.

Gu~\etal\cite{gu2022contrastive} proposed a contrastive domain disentanglement and style augmentation for domain generalization. In particular, domain-style contrastive learning is to properly decompose an image into domain-invariant representation and domain-specific modality representation (\ie, style code), whereas a style augmentation strategy enhances generalizability by combining the randomly generated style codes with given anatomical representation to reconstruct new styles' images.
Meanwhile, Ouyang~\etal\cite{ouyang2020self} devised a superpixel-based SSL with details in pseudo-label generation for few-shot semantic segmentation. By further designing the adaptive local prototype module, they prevent the local information of each class such that it achieves outstanding segmentation performance while improving generalizability. 
Azizi~\etal\cite{azizi2023robust} combines large-scale supervised transfer learning on natural images and intermediate contrastive learning on medical images for specific downstream medical-imaging ML tasks, thereby enhancing the data-efficient generalization performance.

{\REV An emerging and powerful strategy in addressing domain shifts is the combination of pretraining and self-supervised adaptation. This approach leverages the benefits of both large-scale pretraining and task-specific fine-tuning to enhance model generalization.
A notable example is FINE (Feature-level Instance Normalization and Exchange) method proposed by Zhang~\etal~\cite{ZHANG2020116579}, which incorporates the physical model of data generation into the adaptation process. FINE updates the weights of a pretrained network by minimizing a data fidelity loss for each test case, allowing it to better capture features specific to the target domain while maintaining physical consistency.
Similarly, Zhao~\etal~\cite{9253710} introduced Synthetic Multi-Orientation Resolution Enhancement (SMORE). 
SMORE is a self-supervised technique for super-resolution and anti-aliasing of MRI images that does not require external training data.
It works by training a network on high-resolution in-plane slices and applying it to low-resolution through-plane slices to enhance image quality.
}

\paragraph{Adversarial Learning}\label{par:model-adversarial}
Adversarial learning is widely used for learning domain invariant features in machine learning. The key idea of adversarial learning is to introduce adversarial examples during training to make the model more robust to potential attacks or unexpected inputs. These adversarial examples are usually generated by applying minute perturbations to the original input data to deceive the model into making incorrect predictions. 
By incorporating such adversarial examples, the model can better handle real-world scenarios where it may encounter unseen domains, enhancing its ability to make accurate and reliable diagnoses.

Bekkouch~\etal~\cite{bekkouch2021adversarial} proposed the adversarial reconstruction loss to force an encoder to forget style information while extracting useful classification features for hip MRI landmark detection.
Chen~\etal~\cite{chen2020realistic} introduces a realistic adversarial intensity transformation model for data augmentation in MRI that simulates intensity inhomogeneities, common artifacts in MR imaging. This method is a simple yet effective framework based on adversarial training to learn adversarial transformations and to regularize the network for segmentation robustness, which can be used as a plug-in module in general segmentation networks.
Zhang~\etal~\cite{zhang2023domain} proposed an adversarial intensity attack method for medical image segmentation, which exploits an adversarial attack strategy to adjust the intensity distribution in images without altering their content.

\subsubsection{Model Framework}\label{sec:model-framework}
Model framework dives into the architectural and structural strategies deployed to tackle the pervasive challenge of domain shift.
Within this framework, three pivotal approaches are discussed: Ensemble learning, model distillation, and distributed learning.
Together, these strategies represent a comprehensive framework aimed at improving the generalizability of models through innovative architectural solutions and privacy-preserving techniques, ultimately aiming to bridge the gap between diverse medical imaging domains while safeguarding patient privacy.
\paragraph{Ensemble Learning}
Ensemble learning methods are a fundamental approach in machine learning that can significantly enhance model generalization. The key idea behind ensemble models is to build a predictive model by combining the predictions of several base models trained on different subsets of data or using different network architectures. The diverse models can capture varying aspects of unique patterns and feature representation, so their combination could lead to more robust predictions.
In particular, ensemble learning empowers medical imaging systems to achieve robustness and generalization in medical imaging, ultimately contributing to enhanced clinical decision-making and patient care. 

Kamraoui~\etal~\cite{kamraoui2022deeplesionbrain} proposed the Mixture of Calibrated Networks (MCN) for brain tumor segmentation.
The proposed MCN utilizes the complementarity of different base models and takes advantage of their strengths, thus improving the overall system performance.
Specifically, MCN combines the predictions from multiple base models, each with unique calibration characteristics, to deliver more precise tumor boundary definitions and more accurate segmentation results.
Philipp~\etal\cite{philipp2022dynamic} proposed a dynamic CNN for surgical instrument localization, which fuses image and optical flow modalities so that the most reliable information contributes to the prediction. Scalbert~\etal~\cite{scalbert2022test} introduces an ensemble strategy based on multi-domain image-to-image translation for various classification tasks using histology images. Specifically, the proposed method performs image-to-image translation by projecting the target image to the source domains and then ensembles the model prediction of these projected images.

\paragraph{Model Distillation}\label{par:model-distillation}
Model distillation involves transferring the knowledge from a large, sophisticated \textit{teacher} model to a more compact and efficient \textit{student} model.
This process not only preserves the intricate insights and performance capabilities of the teacher model but also ensures that the student model remains lightweight and practical for deployment in environments with stringent computational or memory constraints.

Wang~\etal~\cite{wang2021embracing} proposed Knowledge Distillation for Domain Generalization (KDDG) for the spinal cord gray matter segmentation task.
The authors propose a training strategy that utilizes a gradient filter as a novel regularization term, aiming to simplify the learning task and thereby improve the generalization performance of the model.
The paper articulates that the ''richer dark knowledge``~\cite{hinton2014dark} derived from the teacher network, along with the proposed gradient filter, can significantly mitigate the learning challenge, leading to better generalization in various tasks.
Fernandez-Mart{\'\i}n~\etal~\cite{fernandez2024uninformed} proposed Uninformed Teacher-Student (UTS) for the mitosis localization task, employing a method that distills ``hard'' samples by training a teacher model to identify and retain only the clean, closely matched predictions to annotated mitoses, thus creating a purified training subset.
This subset is used to train a student model, incorporating strong image transformations to challenge and refine the model's focus, enhancing its ability to generalize by learning from a distilled dataset that minimizes noise and irrelevant variability.
\paragraph{Distributed learning}\label{par:model-distributed}
Distributed learning techniques, such as \textit{federated learning} and \textit{privacy preservation}, are essential in medical domain generalization due to patient information's sensitivity in exploiting the data from various institutions~\cite{kaissis2020secure}. When we have used or shared the data from the decentralized device or different data server, privacy concerns may arise in model updates that could leak patients' information~\cite{xu2023federated}. Accordingly, it may violate social ethics that are accompanied by potential risks. To alleviate this fatal issue, advanced techniques such as differential privacy~\cite{abadi2016deep} and federated learning~\cite{chen2020fedhealth} allow models to learn a wide range of data from different institutions or hospitals without directly accessing it while preserving data privacy.

To secure sensitive patient information, Liu~\etal~\cite{liu2021feddg} proposed a privacy-preserving solution with a boundary-oriented episodic learning scheme, which allows us to aggregate model updates from multiple clients without revealing any individual client's data or compromising their privacy. Chen~\etal~\cite{chen2023federated} designed cross-client style transfer using style vectors to improve performance in domain generalization while preserving privacy in federated learning. Meanwhile, Xu~\etal~\cite{xu2023federated} proposed Federated Adversarial Domain Hallucination (FADH), which encodes the information of multiple domains through weight aggregation, as a surrogate for the domain classifier. Using differential privacy, Li~\etal~\cite{li2019privacy} brings sparse vector technique to the patient data owners and only shares intermediate model training updates among them, thus preserving patient data privacy.
By doing so, these approaches ensure the surveillance and security of sensitive patient information while enabling the incorporation of diverse datasets into the learning process, thereby promoting model generalization.

\subsubsection{Other Model-based DG}\label{sec:model-others}

\paragraph{Geometric learning}
Geometric learning~\cite{bronstein2021geometric} is an approach that leverages the intrinsic geometric structure of data, often residing in non-Euclidean spaces.
Non-Euclidean spaces refer to geometric environments that do not adhere to Euclidean geometry, \eg, graphs, topologies, and manifolds, often encountered in \MIA.
Here, geometric learning harnesses these intrinsic data geometries, exploiting the geometric information to better generalize across different domains.
By modeling the complex correlations in high-dimensional data, geometric learning can better handle irregularities inherent in medical imaging. 
Readers are referred to~\cite{li2022out} for a comprehensive review of the OOD generalization on graphs.
The following paragraph explores the geometric learning techniques specifically proposed for \MIA tasks.

Nguyen~\etal~\cite{nguyen2023lvm}  developed a graph-matching algorithm employing a $k$-nearest neighbors approach to construct graphs from medical images, where nodes represent distinctive image regions and edges define spatial relationships. Their Learned Vertex Matching (LVM) method analyzes structural similarities and differences across images, enhancing abnormality detection and segmentation tasks.
In a similar approach, Seenivasan~\etal~\cite{seenivasan2022biomimetic} proposed a graph network for surgical scene understanding.
They utilized graph learning to understand interactions in the surgical scene by embedding the visual and semantic features of instruments and tissues into graph nodes. 
Santhirasekaram~\etal~\cite{santhirasekaram2023topology} proposed a hierarchical topology preservation method for medical image segmentation tasks.
Their method constrains the latent space of a deep learning model to a dictionary of base components, which are chosen to capture the limited structural variability found across patients' medical images.
This dictionary is learned through vector quantization, and a topological prior is incorporated into the sampling process using persistent homology, which ensures topologically accurate segmentation maps.
\input{tables/analysis_level_dg}
\paragraph{Distributionally Robust Optimization\label{sec:model-optimization}}
Distributionally Robust Optimization (DRO)~\cite{delage2010distributionally} is a model-level domain generalization method aiming to optimize model performance over the worst-case distribution within a specified uncertainty set. In other words, instead of optimizing the model's performance based on a single training data distribution, DRO tries to ensure good performance across a range of possible data distributions.

Bissoto~\etal~\cite{bissoto2022artifact} utilized the Group Distributionally Robust Optimization (GDRO)~\cite{sagawa2019distributionally} in their skin lesion classification model.
GDRO extends the DRO framework by considering groups, or ``environments'', in the data distribution.
They partitioned the training data into different environments based on the presence of various artifacts, such as hair, ruler marks, and dark corners.
These environments were then used to train the model under the GDRO framework.
Goel~\etal~\cite{goel2020model} enhanced Generalized Distributionally Robust Optimization (GDRO) by introducing class-conditional Subgroup DRO (SGDRO) for skin lesion classification, which refines risk minimization by considering both broad data groups and more granular subgroups defined by class-specific traits. SGDRO optimizes for the worst-case scenario within each subgroup across different environments, resulting in a model that is better equipped to handle complex data distributions and more resistant to distributional shifts.

%% file: tables/model_level_dg.tex
\begin{table*}[htbp]\scriptsize\setlength{\tabcolsep}{7.pt}
    \caption{Model-level domain generalization methods. Methods categorized by different settings for source and target domains, task, organ, and modality.}
    \label{tab:model-level}
    \centering
    \begin{tabular}{C{2.4cm}C{1.5cm}C{0.6cm}C{1.5cm}C{1.8cm}C{1.8cm}C{1.8cm}C{1.8cm}}
    \toprule
     {\bfseries Method} & {\bfseries Specific} & {\bfseries Ref.} & {\bfseries Source} & {\bfseries Target} & {\bfseries Task} & {\bfseries Organ} & {\bfseries Modality}\\
\toprule
    Learning Strategy & \mrow{1}{Meta-learning} & \cite{khandelwal2020domain} & Multi & Site & Segmentation & Spinal &CT\lc{f:csi,f:xvertseg,f:verse}  \\\ccmrule
     &  & \cite{dou2019domain} & Multiple & Site & Segmentation & Brain & MRI \\\ccmrule
        & & \cite{liu2020shape} & Multiple & Site & Segmentation & Prostate & MRI\lc{f:promise,f:i2cvb,f:nci-isbi} \\\ccmrule
    & & \cite{liu2021semi} & Multiple & Site, Sequence & Segmentation & Cardiac\lc{f:mms}, Spinal\lc{f:scgm} & MRI \\\ccmrule
    & & \cite{hu2023map} & Multi & Site & Segmentation & Retinal & Fundus\lc{f:drive,f:stare,f:aria,f:prime-fp20}, OCT\lc{f:rose,f:octa-500}, FC\lc{f:recovery-fa19} \\\ccmrule
    & & \cite{10251660} & Single, Multiple & Site & Classification & Brain & Functional MRI\lc{f:abide} \\
         \cmrule
    & \mrow{1}{Self-supervised Learning} & \cite{gu2022contrastive} & Multiple & Site & Segmentation & Retinal &  Fundus\lc{f:drishti-gs,f:rim-one-r3,f:refuge}   \\\ccmrule
    &  & \cite{gu2022cddsa} & Multiple & Site & Segmentation & Retinal &  Fundus\lc{f:drishti-gs,f:rim-one-r3,f:refuge} \\\ccmrule
    & & \cite{ouyang2020self} & Multiple & Modality & Segmentation & Abdominal\lc{f:sabsct,f:chaos}, Cardiac\lc{f:ms-cmrseg} & CT, MRI\\\ccmrule
    & & \cite{azizi2023robust} & Multiple & Site & Classification & Skin, Retinal, Chest\lc{f:mimic-cxr,f:chestx-ray14,f:chexpert}, Breast\lc{f:camelyon17} & Histology, Fundus, X-ray, Mammography\\\ccmrule
    & & \cite{hu2023devil} & Single & Site & Segmentation & Retinal & Fundus\lc{f:riga+} \\
         \cmrule
    & \mrow{1}{Adversarial Learning} & \cite{bekkouch2021adversarial} & Single, Multiple & Sequence & Detection & Skin, Hip & MRI\lc{f:ham10000} \\\ccmrule
    & & \cite{chen2020realistic} & Multiple & Site & Segmentation & Cardiac & MRI\lc{f:acdc} \\\ccmrule
    & & \cite{zhang2023domain} & Multiple & Site & Segmentation & Retinal\lc{f:drishti-gs,f:rim-one-r3,f:refuge}, Prostate\lc{f:promise,f:i2cvb,f:nci-isbi} & Fundus, MRI \\\ccmrule
    & & \cite{cheng2023adversarial} & Single & Site & Classification & Breast & Histology\lc{f:camelyon17} \\
    \midrule
    
    Model Framework &\mrow{1}{Ensemble Learning}& \cite{kamraoui2022deeplesionbrain} & Multiple & Sequence, Site & Segmentation & Brain & MRI\lc{f:msseg,f:isbi-ms} \\\ccmrule
    & & \cite{philipp2022dynamic} & Single & Sequence & Localization & Surgical Scene & Video Frames\lc{f:surgicalactions160,f:cataract-101} \\\ccmrule
    & & \cite{scalbert2022test} & Single, Multiple & Site & Classification & Colon\lc{f:kather16,f:crc-tp,f:kather18}, Breast\lc{f:camelyon17} & Histology\\
         \cmrule         
    &\mrow{1}{Model Distillation}& \cite{wang2021embracing} & Single & Site & Segmentation & Spinal & MRI\lc{f:scgm}  \\\ccmrule
    & & \cite{fernandez2024uninformed} & Multi & Site & Detection & Breast & Histology\lc{f:midog,f:tupac,f:mitos,f:ccmct} \\\ccmrule
    & & \cite{galappaththige2024generalizing} & Single, Multi & Site & Classification & Retinal & Fundus\lc{f:kdr,f:aptos,f:messidor} \\\ccmrule
    & & \cite{wang2024leveraging} & Single& Site& Segmentation& Prostate & MRI \\\ccmrule
    & & \cite{gao2023desam} & Single& Site& Segmentation& Prostate & MRI\lc{f:promise,f:i2cvb,f:nci-isbi} \\\ccmrule
    & & \cite{chen2023ma} & Multi & Site & Segmentation & Abdominal, Prostate, Surgical Scene & CT\lc{f:sabsct,f:msd}, MRI\lc{f:promise,f:i2cvb,f:nci-isbi}, Video\lc{f:endovis-robot} \\
              \cmrule

     &  \mrow{1}{Distributed Learning} & \cite{liu2021feddg} & Multiple & Site & Segmentation & Retinal\lc{f:rim-one-r3,f:refuge}, Prostate\lc{f:promise,f:nci-isbi} & Fundus, MRI \\\ccmrule
     &  & \cite{chen2023federated} & Multiple & Site & Classification & Breast & Histology\lc{f:camelyon17} \\\ccmrule
     &  & \cite{xu2023federated} & Multiple & Site & Segmentation & Retinal & Fundus\lc{f:refuge} \\\ccmrule
     &  & \cite{li2019privacy} & Multiple & Modality & Segmentation & Tumor & MRI\lc{f:brats} \\
    \midrule
    Other & \mrow{1}{Geometric Learning} & \cite{nguyen2023lvm} & Multiple & Modal, Sequence, Site & Classification, segmentation, detection & \multicolumn{2}{c}{Multiple organs, modalities from 55 datasets}  \\\ccmrule
     &  & \cite{seenivasan2022biomimetic} & Multiple & Site & Segmentation & Surgical Scene & Video Frames\lc{f:endovis-robot} \\\ccmrule
     &  & \cite{santhirasekaram2023topology} & Single & Site & Segmentation & Abdominal\lc{f:sabsct}, Cardiac\lc{f:mms}, Prostate\lc{f:nci-isbi} & MRI, CT \\\cmrule
    &\mrow{1}{Distributionally Robust Optimization} & \cite{bissoto2022artifact} & Multiple & Site & Classification & Skin & Dermatology\lc{f:isic} \\\ccmrule
    & & \cite{goel2020model} & Multiple & Site & Classification & Skin & Dermatology\lc{f:isic} \\
\bottomrule
\end{tabular}
\end{table*}

%% file: tables/analysis_level_dg.tex
\begin{table*}[htbp]\scriptsize\setlength{\tabcolsep}{7.pt}
    \caption{Analysis-level domain generalization methods. Methods categorized by different settings for source and target domains (see \secref{sec:settings}), task, organ, and modality used in experiment.}
    \label{tab:analysis-level}
    \centering
    \begin{tabular}{C{2.2cm}C{0.6cm}C{1.5cm}C{1.8cm}C{1.8cm}C{1.8cm}C{1.8cm}}
    \toprule
     {\bfseries Method} & {\bfseries Ref.} & {\bfseries Source} & {\bfseries Target} & {\bfseries Task} & {\bfseries Organ} & {\bfseries Modality}\\
\toprule
     Interpretable AI & \cite{wang2020proactive} & Single & Site & Classification, Segmentation & Retinal\lc{f:a2asdoct}, Cardiac\lc{f:lidc-idri} & OCT, CT \\\ccmruleaa
     & \cite{karim2021deepkneeexplainer} & Single & Site & Classification, Segmentation & Knee & MRI\lc{f:most} \\\ccmruleaa
     & \cite{wang2022seeg} & Multiple & Site & Classification & Brain & EEG\lc{f:mayo,f:fnusa} \\\ccmruleaa
     & \cite{fan2022invnorm} & Multiple & Sequence & Detection & Gastrointestinal & Endoscopy \\
    \midrule
     Transferability & \cite{gao2023bayeseg} & Single & Sequence, Site & Segmentation & Cardiac\lc{f:emidec,f:acdc,f:mm-whs,f:ms-cmrseg}, Prostate\lc{f:promise,f:i2cvb,f:nci-isbi} & MRI, CT\\\ccmruleaa
     & \cite{yuan2022not} & Single, Multiple & Site & Classification & Breast & Histology\lc{f:camelyon17} \\
    \midrule
     Causality & \cite{mahajan2021domain} & Single, Multiple & Sequence & Classification & Chest & X-ray\lc{f:rsna-pd,f:chexpert,f:cxr8} \\\ccmruleaa
     & \cite{wang2021harmonization} & Multiple & Site & Classification & Brain & MRI\lc{f:istaging,f:adni} \\\ccmruleaa
     & \cite{ouyang2022causality} & Single & Modality, Sequence, Site & Segmentation & Prostate\lc{f:promise,f:i2cvb,f:nci-isbi}, Abdominal\lc{f:sabsct,f:chaos}, Cardiac\lc{f:ms-cmrseg} & MRI, CT\\
\bottomrule
\end{tabular}
\end{table*}

%% file: sections/methods/analysis.tex
\subsection{Analysis-level Generalization}\label{sec:analysis-level}
{\REV Analysis-level DG refers to techniques that focus on understanding and interpreting the behavior of domain-generalized models.
These methods aim to: (1) provide insights into how models make decisions across different domains, (2) evaluate the extent of a model's generalization, (3) identify potential biases or failure modes in generalized models, and (4) enable trust and adoption of DG models in clinical settings.
Unlike data-level, feature-level, or model-level approaches that primarily aim to improve generalization performance, analysis-level methods are concerned with the post-hoc examination and interpretation of already generalized models. This is particularly crucial in \MIA, where understanding model decisions is essential for clinical validation and trust.

Analysis-level DG techniques face unique challenges, as they must provide interpretations that are consistent and meaningful across multiple domains, often with varying characteristics. These methods must balance the trade-off between model performance and interpretability, especially for complex, highly generalizable models.}

\subsubsection{Interpretable AI\label{sec:analysis-interpretable}}
Interpretable AI aims to develop techniques that help evaluate and debug a model's decisions making process.
Interpreting domain generalization models is more challenging as these models have special architectures and learning paradigms to accommodate the novel DG settings (\eg, cross-modality).
Hence, interpretable AI for DG proposes new techniques to visualize the model's output given heterogeneous data, such as multi-modal~\cite{karim2021deepkneeexplainer} and temporal~\cite{wang2022seeg} data.
In \MIA, AI's ability to adapt to new, unseen data from various hospitals or demographic backgrounds is crucial for diagnosing and determining treatment paths accurately.
Interpretable AI, especially under DG setting, is essential as it allows healthcare professionals to understand and trust AI's decisions, thereby enhancing patient care and safety through transparency and clinical evidence validation.
For a general overview of interpretable AI for \MIA, readers are referred to the survey by Singh~\etal~\cite{singh2020explainable} and Van~\etal~\cite{van2022explainable}.
In the following paragraphs, we present domain generalization techniques for interpretable AI specifically designed for \MIA.

Dong~\etal~\cite{wang2020proactive} proposed a saliency map-based method for lung lesion classification that uses a contrastive learning scheme incorporating synthetic causal interventions. This technique utilizes weighted backpropagation to generate a saliency map that visualizes and highlights causally relevant areas in the data, thereby improving our understanding of the model's decision-making process.
Karim~\etal~\cite{karim2021deepkneeexplainer} proposed DeepKneeExplainer, a CAM-based interpretable AI method for multimodal knee osteoarthritis diagnosis.
The DeepKneeExplainer uses an explainable neural ensemble method to improve performance by implicitly reducing the generalization error and using CAM to visualize the model's decision.
Similarly, Wang~\etal~\cite{wang2022seeg} proposed a novel focal domain generalization loss and used Grad-CAM++~\cite{chattopadhay2018grad} to visualize the pathological activity from stereo-electroencephalogram (sEEG). 

\paragraph{Transferability}\label{sec:analysis-transferability}
Methods that provide interpretability in one domain may not necessarily transfer well to other domains.
Since different medical imaging data and tasks may require different interpretability approaches, ensuring that interpretability methods can be effectively applied across diverse domains is challenging.
Gao~\etal~\cite{gao2023bayeseg} proposed BayeSeg for interpretable medical image segmentation.
One of the key advantages of BayeSeg is its ability to control the performance and interpretability tradeoff.
By approximating the posterior distributions of the shape, appearance, and segmentation, BayeSeg captures the statistical relationships between these variables.
This statistical modeling allows users to adjust the weights of the variational loss terms in BayeSeg to prioritize different aspects of the segmentation process, allowing them to control the tradeoff between interpretability and performance.
Yuan~\etal~\cite{yuan2022not} proposed a method for augmenting histopathology images using text prompts (\eg, ``synthesize image of a lymph node in the style of S*'').
To tackle the challenge of transferability, authors proposed to leverage text-to-image (T2I) generators as a means of enabling interpretable interventions for robust representations.
The authors argue that T2I generators offer unprecedented capability and flexibility in approximating image interventions conditioned on natural language prompts.
By using T2I generators, the proposed method can provide a more interpretable and domain-agnostic approach that can be effectively applied across diverse domains.

\subsubsection{Causalility}\label{sec:analysis-causality}
Causality refers to the relationship between variables in a causal system, where one variable (the cause) directly affects or influences another variable (the effect).
In domain generalization, causality focuses on understanding the underlying causal mechanisms that lead to the differences between source and target domains.
It aims to identify the causal factors invariant across different domains and responsible for the targeted \MIA task.
By understanding and leveraging causality, domain generalization methods can effectively generalize the learned knowledge from a source domain to target domains with different distributions.
Readers are referred to a survey by Seth~\etal~\cite{sheth2022domain} for a deeper insight into the causal perspective of domain generalization for general tasks.
In the following paragraph, we explore several approaches of causal learning specifically designed for domain generalization for \MIA tasks.

Mahajan~\etal~\cite{mahajan2021domain} proposed a causality-aware domain generalization method for pneumonia detection using chest X-ray images.
They used a causal Bayesian network to model the relationships among the domain, the image features, and the class label.
By explicitly modeling the causal relationships, they were able to identify the common causal features that are invariant across domains and are important for predicting the presence of pneumonia.
Wang~\etal~\cite{wang2021harmonization} used a causal graph-based approach for Alzheimer’s disease diagnosis using MRI.
They modeled the causal relationships among imaging sites, gender, age, and imaging features using a Structural Causal Model (SCM).
By performing counterfactual inference on the model, they could generate harmonized data that simulate the imaging data as if it came from the same site.
This approach effectively removed the site-specific confounding factors and improved the generalization of the trained model across different sites.
Similarly, Ouyang~\etal~\cite{ouyang2022causality} proposed a causal learning framework for single-source domain generalization in CT image segmentation.
They introduced a SCM to represent the causal relationships between the input data, the domain shift variables, and the task-specific output.
The SCM allows for the identification of invariant causal factors shared across different domains, which can be used to improve the generalization of the models.

%% file: sections/methods/others.tex
\section{DG Under Limited Source}\label{sec:source-limted-DG}
\begin{table}[h]\scriptsize
\centering
\caption{Comparison of Source-limted DG}
\label{tab:DG_Comparison_Notation}
\begin{tabular}{lc}
\toprule
\textbf{DG Paradigm} & Access to Source Domain \\ \toprule
Multi-source DG & Full access (\ie, $\{\mathcal{S}^i\}_{i=1}^M$)\\\arrayrulecolor{lightgray!20}\midrule
Single-source DG & Single source domain (\ie, $\mathcal{S}^1$) \\\midrule
Unsupervised DG & Unlabeled source domain \\\midrule
Open-set DG & With concept (label) shift \\\midrule
Source-free DG & Pretrained source model only \\\midrule\arrayrulecolor{black}
Zero-shot DG & Auxiliary information only \\ \bottomrule
\end{tabular}
\end{table}

In this section, we examine the challenge of domain generalization under severe restrictions in the source domain, as detailed in Table~\ref{tab:DG_Comparison_Notation}. These particular DG scenarios have received relatively less attention in the field of \MIA due to their extreme conditions.
Our review aims to shed light on these under-explored areas and their implications for \MIA.

\subsection{Single-source Domain Generalization}
Single-source domain generalization (SSDG) assumes that there is only one source domain to learn from.
Due to the lack of diversity in the training data, most SSDG propose augmentation-based solutions, both in data and feature space, to simulate a broader range of domain variability.
For example, augmentation-based SSDG inlcude color normalization~\cite{zhang2022semi} (\secref{par:image-processing}), frequency-based~\cite{li2023frequency} (\secref{par:surrogate}), randomization~\cite{zhang2022robust,gomathi2022novel} (\secref{par:data-randomization}), generative models~\cite{yang2023learning} (\secref{par:data-generative}), feature augmentation~\cite{wen2024denoising} (\secref{sec:feat_aug}), contrastive learning~\cite{hu2023devil} (\secref{par:model-ssl}), adversarial learning~\cite{xu2022adversarial,cheng2023adversarial} (\secref{par:model-adversarial}).
Recent \textit{bleeding-edge} SSDG include transferring knowledge from large-scale pretrained models, \ie variations of model distillation~\cite{galappaththige2024generalizing,wang2024leveraging,gao2023desam} (\secref{par:model-distillation}).

\subsection{Open-set Domain Generalization}
\input{tables/source_limited_dg}
Open-set domain generalization (OSDG) refers to DG techniques that specialize in capturing and correcting the concept shift in addition to the covariate shift (\secref{sec:domain-shift}).
Yang~\etal~\cite{yang2022full} proposed a simple feature-based semantics score function to consider both detecting label shift and being tolerant to covariate shift as in-distribution. Mahajan~\etal~\cite{mahajan2021connection} investigated the theoretical relationship of whether better OOD generalization leads to better privacy for ML models in practice and showed that capturing stable features from models represents superior open-set generalization with robustness. 
Zheng~\etal~\cite{zheng2023single} proposed Open-Set Single-Domain Generalization for Multiple Cross-Matching (MCM) for the open-set lung cancer diagnosis.
This work delves into an open-set single-source DG problem where the source domain only contains data with unique class names, while the target domain contains multiple unseen class names.
Puli~\etal~\cite{puli2021out,puli2022nuisances} and Gao~\etal~\cite{gao2022out} deal with the spurious correlation or variations underlying several confounding variables in terms of causal perspective to circumvent the open-set problem.

\subsection{Other under-explored DGs}
In this section, we introduce some under-explored DGs where only preliminary research has been done in the field of \MIA.
\parggg{Unsupervised domain generalization}
refers to methods that enable a model to learn useful, domain-invariant features from unlabeled source data such that it can perform well on unseen domains.
\parggg{Source-free domain generalization}
places an extreme constraint on privacy-preserving models (\secref{par:model-distributed}) where it assumes source data is inaccessible, but only the pretrained source model is available.
\parggg{Zero-shot domain generalization}
is another extreme case of DG where only the auxiliary information (\eg, meta-data) from the source domain is available.
A possible solution~\cite{galappaththige2024generalizing,wang2024leveraging,gao2023desam,chen2023ma} (\secref{par:model-distillation}) to these extreme cases of DGs involves a pretrained foundation model combined with zero-shot prompting techniques, though these topics fall beyond the scope of this survey due to the lack of literature in this topic.

%% file: tables/source_limited_dg.tex
\begin{table}[ht]
    \caption{Open-set domain generalization methods. Methods categorized by different settings for source and target domains (see \secref{sec:settings}), task, organ, and modality used in experiment.}
    \label{tab:source-limited}
    \centering
    \resizebox{1\linewidth}{!}{
    \begin{tabular}{ccccccc}
    \toprule
     {\bfseries Method} & {\bfseries Ref.} & {\bfseries Source} & {\bfseries Target} & {\bfseries Task} & {\bfseries Organ} & {\bfseries Modality}\\
\toprule
     \mrow{9.2}{Open-set DG} & \cite{yang2022full} & Single & Site & Classification & Chest & X-ray\lc{f:covid-ct,f:hannover-cv,f:rsna-ba,f:bimcv,f:actualmed} \\\ccmruleaa
      & \cite{mahajan2021connection} & Multiple & Site & Classification & Chest & X-ray\lc{f:chexpert} \\\ccmruleaa
      & \cite{zheng2023single} & Single & Site & Classification & Chest & X-ray\lc{f:bimcv} \\\ccmruleaa
      & \cite{puli2021out} & Single & Site & Classification & Chest & X-ray\lc{f:mimic-cxr,f:chexpert} \\\ccmruleaa
      & \cite{puli2022nuisances} & Single & Site & Classification & Chest & X-ray\lc{f:mimic-cxr,f:chexpert} \\\ccmruleaa
      & \cite{gao2022out} & Multiple & Site & Classification & Breast & Histology\lc{f:camelyon17} \\
\bottomrule
\end{tabular}
}
\end{table}

%% file: sections/insights.tex
\section{Future Directions}\label{sec:future}
\input{figures/tikz/suggestion}

\input{tables/critical_analysis}
Domain generalization for \MIA is a rapidly evolving field with several promising directions for future research.
In this section, we outline some important areas that warrant further exploration.

\subsection{Source-limited Domain Generalization}
Unlike the common focus on scenarios involving multiple sources, there is a significant gap in research for methods tailored to domain generalization under limited source (\secref{sec:source-limted-DG}).
For example, single-source domain generalization requires the model to generalize well to unseen domains even when only a single source domain is available for training. This scenario often arises in medical imaging, where data collection can be resource-intensive, and privacy concerns may limit access to multiple sources. Future research should explore novel methods that effectively address the challenges of source-limited domain generalization, such as robustness to concept shift in the presence of covariate shift.

\subsection{Medical Foundation Model}
Foundation models~\cite{Bommasani2021FoundationModels} refer to hyper-scale models that has been trained on massive and diverse datasets.
Arguably, this new class of model is a step towards next-level artificial intelligence that has state-of-the-art zero-shot generalizability performances.
However, development of foundation model for \MIA is challenging as medical datasets are heterogeneous and hard to collect in large scale.
Despite these challenges, research on medical foundation models is very active and shows great potential~\cite{moor2023foundation}.
Recently, Segment Anything Model (SAM)~\cite{gao2023desam,wang2024leveraging} has shown promising results for various source-limited DG in segmentation tasks for \MIA.

\subsection{Benchmark Datasets}\label{sec:insight-dataset}
Many DG methods for \MIA rely on custom datasets that mix private and public datasets, treating each dataset as a distinct domain. 
The reproducibility of these custom datasets is extremely low because the processes for data splitting, preprocessing, and annotation vary widely and are hard to duplicate.
Therefore, there is a need to establish standardized benchmark datasets that reflect the diversity and challenges encountered in real-world medical imaging scenarios.
These benchmark datasets should cover various imaging modalities, patient populations, and imaging protocols to facilitate the fair and rigorous evaluation of domain generalization methods.
Additionally, benchmark datasets for different settings of domain generalization, \ie, multi-source, single-source, cross-site, cross-sequence, cross-modality, covariate shift, and concept shift, should be developed to enable comprehensive evaluation and comparison of different techniques.
{\REV Readers are referred to the Supplementary and the interactive website (\url{https://milab.korea.ac.kr/dg-dataset}) for a review of existing DG benchmark datasets and a summary of public datasets.}

\subsection{Suggestions for Domain Generalization for \MIA}\label{sec:strategy}
There exist a number of empirical evaluations of commonly used domain generalization techniques suggesting appropriate methods for specific tasks at hand. For example, Korevaar~\etal~\cite{korevaar2023failure} evaluated methods on benchmark datasets, whereas Zhang~\etal~\cite{zhang2021empirical} and Galappaththige~\etal~\cite{galappaththige2024generalizing} evaluated them on some custom cross-site datasets consisting of publicly available datasets.
While these benchmarks shed light on the capabilities of certain techniques in domain generalization, they do not offer a comprehensive guide for problem-specific and task-specific strategies throughout the \MIA workflow.
To this end, we critically analyze different DG methods (Table~\ref{tab:critical}) and suggest strategies to incorporate domain generalization into the model (Fig.~\ref{fig:suggestion}).

%% file: figures/tikz/suggestion.tex
\begin{figure}[t]
    \centering
    \caption{Problem-specific suggestion for strategies for integrating domain generalization into \MIA workflow. Diamond box indicates the start terminator, angled boxes indicate the process, and round boxes indicate the decision.}
    \vspace{0.3cm}

\definecolor{black}{RGB}{0, 0, 0}
\definecolor{gray}{RGB}{51, 51, 51}
\definecolor{purple}{RGB}{175, 179, 255}

\definecolor{diamond-box-color}{RGB}{130, 160, 160}
\definecolor{angled-box-color}{RGB}{125, 145, 160}
\definecolor{rounded-box-color}{RGB}{110, 130, 160}

\tikzstyle{rectangle_round} = [rectangle, text width=1.77cm, rounded corners, minimum width=1.23cm, minimum height=1.33cm, text centered, font=\footnotesize, color=gray, draw=black, line width=0.1, fill=teal!10]
\tikzstyle{rectangle_angled} = [rectangle, text width=1.77cm, minimum width=1.23cm, minimum height=1.33cm, text centered, font=\footnotesize, color=gray, draw=black, line width=0.1, fill=teal!20]
\tikzstyle{diamond_box} = [diamond, text width=0.75cm, minimum width=0.75cm, minimum height=0.1cm, text centered, font=\footnotesize, color=gray, draw=black, line width=0.1, fill=teal!30]

\begin{tikzpicture}[node distance=3cm]
\node (frequency) [rectangle_round] {\secref{par:frequency}\\Frequency-based};
\node (input) [diamond_box, right of = frequency] {Start};
\node (related) [rectangle_round , right of=input] {\secref{sec:related}\\Related ML tasks};

\node (sufficient) [rectangle_angled , below of=frequency, yshift=1cm] {Sufficient data?};
\node (raw) [rectangle_angled , right of=sufficient] {Raw measurement available?};
\node (access) [rectangle_angled , right of=raw] {Access to $\mathcal{S}_{target}$?};

\node (augmentation) [rectangle_round , below of=sufficient, yshift=1cm] {\secref{sec:data-augmentation}\\Data\\Augmentation};
\node (preprocessing) [rectangle_angled , right of=augmentation] {Preprocessing required?};
\node (specifictask) [rectangle_angled , right of=preprocessing] {Specific task or problem at hand?};

\node (analysis) [rectangle_angled , below of=augmentation, yshift=1cm] {Analysis or interpretability required?};
\node (problem) [rectangle_round , right of=analysis] {\secref{sec:data-problem-specific}\\Problem-specific};
\node (imageprocessing) [rectangle_round , right of=problem] {\secref{par:image-processing}\\Image Processing};

\node (analysislevel) [rectangle_round , below of=analysis, yshift=1cm] {\secref{sec:analysis-level}\\Analysis-level DG};
\node (specialalgorithm) [rectangle_angled , right of=analysislevel] {Able to use special algorithm or architecture?};
\node (modellevel) [rectangle_round , right of=specialalgorithm] {\secref{sec:model-level}\\Model-level DG};

\node (alignment) [rectangle_round , below of=analysislevel, yshift=1cm] {\secref{sec:feature-alignment}\\Feature Alignment};
\node (invariance) [rectangle_angled , right of=alignment] {Invariant \textit{v.s.} Disentangled Features};
\node (disentanglement) [rectangle_round , right of=invariance] {\secref{sec:feature-disentanglement}\\Feature Disentanglement};

\draw[-Stealth] (input) -- (access);
\draw[-Stealth] (access) -- node[anchor=west] {Y} (related);
\draw[-Stealth] (access) -- node[anchor=north] {N} (raw);
\draw[-Stealth] (raw) -- node[anchor=north] {Y} (frequency);
\draw[-Stealth] (raw) -- node[anchor=north] {N} (sufficient);
\draw[-Stealth] (sufficient) -- node[anchor=north] {Y} (preprocessing);
\draw[-Stealth] (sufficient) -- node[anchor=west] {N} (augmentation);
\draw[-Stealth] (preprocessing) -- node[anchor=north] {Y} (specifictask);
\draw[-Stealth] (specifictask) -- node[anchor=north] {Y} (problem);
\draw[-Stealth] (specifictask) -- node[anchor=west] {N} (imageprocessing);

\draw[-Stealth] (preprocessing) -- node[anchor=north] {N} (analysis);
\draw[-Stealth] (analysis) -- node[anchor=west] {Y} (analysislevel);
\draw[-Stealth] (analysis) -- node[anchor=north] {N} (specialalgorithm);
\draw[-Stealth] (specialalgorithm) -- node[anchor=north] {Y} (modellevel);
\draw[-Stealth] (specialalgorithm) -- node[anchor=west] {N} (invariance);
\draw[-Stealth] (invariance) -- node[anchor=north] {Inv.} (alignment);
\draw[-Stealth] (invariance) -- node[anchor=north] {Dis.} (disentanglement);

\end{tikzpicture}    
    \label{fig:suggestion}
\end{figure}

%% file: tables/critical_analysis.tex
\begin{table*}[htbp]\scriptsize
\caption{Critical Analysis of Domain Generalization Methods\label{tab:critical}}
\begin{tabular}{C{1cm}C{1cm}C{2.4cm}p{6cm}p{6cm}}
    \toprule
     {\bfseries Level} &{\bfseries Method} & {\bfseries Specific} & {\bfseries Strengths} & {\bfseries Limitations}\\
    \bottomrule

Data & Image Process. & Intensity Normalization & Uniformity across images; improved ML performance & May reduce contrast; relies on similarity assumptions \\\cmrulemethods
 &  & Histogram Matching & Adapts to reference histograms improving consistency & Depends on reference choice; less suitable for multi-domain \\\cmrulemethods
 &  & Color Normalization & Standardizes color for consistency and feature recognition & Could alter diagnostic features; mostly for stain images \\\cmrulemethodss
 & Surrogate & Frequency-based & Separates amplitude/phase for style/content manipulation & Sensitive to noise; limited by domain applicability \\\cmrulemethods
 &  & Raw Signals & Manages early-stage data manipulation; captures latent info & Complexity/accessibility issues; limited applicability \\\cmrulemethods
 &  & Dictionary Learning & Captures sparse representation and structural similarities & Computationally intense; may not handle unique features \\\cmrulemethodss
 & Augment & Randomization & Increases data diversity and simplicity & Limited augmentation scope; may create unrealistic images \\\cmrulemethods
 &  & Adversarial  & Improves robustness through targeted examples & Computationally intense; risks catastrophic forgetting \\\cmrulemethods
 &  & Generative & Generates diverse and realistic data & Complex to train; prone to modal-collapse \\\cmrulemethodss
 & Problem-specific & Cross-modal & Enhances unpaired data utilization & Complex and relies on synthetic data quality \\\cmrulemethods
 &  & Stain Normalization & Enhances histopathological analysis & Limited applicability; risks of over-normalization \\\cmrulemethodsss
Feature & Feature Align. & Feature Normalization & Standardizes statistical distributions efficiently & Depends on choice of domain-specific statistics \\\cmrulemethods
 &  & Dissimilarity-based & Mitigates domain shift through direct measures & Computationally complex and challenges in metric selection \\\cmrulemethodss
 & Implicit Disent. & Information Theoretic  & Enhances interpretability and adapts to complex structures & Difficult to estimate MI; depends on estimation quality \\\cmrulemethods
 &  & Contrastive  & Improves sample efficiency; robustness to domain shifts & Depends on pair quality; risks representational collapse \\\cmrulemethods
 &  & Variational  & Models uncertainty; direct latent space regularization & Balance between fidelity and disentanglement \\\cmrulemethodss
 & Explicit Disent. & Conditional Representation & Enhances contextual adaptation; targeted feature learning & Complexity leading to potential overfitting on condition \\\cmrulemethods
 &  & Feature Regularization & Robustness to variations; prevents over-reliance on features & Sensitive to choice of  hyperparameters\\\cmrulemethodss
 & Others & Feature Augmentation & Incorporates domain knowledge within latent space & Complex and can overfit to augmented features \\\cmrulemethods
 &  & Kernel-based Learning & Incorporates domain knowledge efficiently & Selection of kernel is critical; scalability issues \\\cmrulemethodsss
Model & Learning Strategy & Meta-learning & Enables rapid adaptation and efficient learning & Faces complex optimization; overfitting to meta-tasks \\\cmrulemethods
 &  & Self-supervised Learning& Utilizes unlabeled data for enhanced features & Dependent on pretext task design\\\cmrulemethods
 &  & Adversarial Learning& Learns domain-invariant features and enhances robustness & Vulnerable to adversarial attacks; computationally complex \\\cmrulemethodss
 & Model Framework & Ensemble Learning & Increases robustness and leverages model diversity & High computational cost and implementation complexity \\\cmrulemethods
 &  & Distillation & Highly efficient; preserves ID performance & Dependent on choice of teacher; complex training process \\\cmrulemethods
 &  & Distributed Learning& Preserves privacy and improves scalability & Faces communication overhead and heterogeneity issues \\\cmrulemethodss
 & Others & Geometric & Handles non-Euclidean data well & High complexity; limited applicability \\\cmrulemethods
 &  & DRO & Optimizes across worst-case scenarios for robustness & Complex in defining uncertainty sets\\\cmrulemethodsss
Analysis &  & Interpretable AI & Enhances trust; model debugging & Interpretability-complexity tradeoff \\\cmrulemethods
 &  & Causality & Focuses on invariant features for robust generalization & Complex model development and limited scalability\\
\bottomrule
\end{tabular}
\end{table*}

%% file: sections/conclusion.tex
\section{Conclusion}
Domain generalization is a crucial capability for modern medical image analysis, aimed at creating machine learning models capable of handling a wide variety of data distributions arising from variations in, for example, imaging protocols, patient demographics, and equipment. This paper provides a comprehensive review of domain generalization techniques, extending beyond the methodological hierarchy of previous surveys and considering the implications of domain generalization on the entire \MIA workflow. Our focus includes every stage of the decision-making process, from data acquisition to pre-processing, prediction, and analysis.

We have also highlighted and discussed the current benchmark datasets, emphasizing the necessity to expand the spectrum of these resources.
Moreover, we shed light on potential directions for future research in this field.
While domain generalization for \MIA is still a rapidly evolving field, it is clear that it holds significant promise for improving patient care by enhancing the robustness and reliability of \MIA models.
As we continue to address the challenges and capitalize on the opportunities, we anticipate seeing substantial improvements in the efficiency and accuracy of \MIA workflows, leading to more personalized and effective patient treatments.
This, in turn, will help the healthcare sector move towards the broader goal of precision medicine, providing each patient with care that is uniquely tailored to their health profile.

%% file: biography.tex
\vfill\null
\begin{IEEEbiography}[{\includegraphics[width=1in,clip,keepaspectratio]{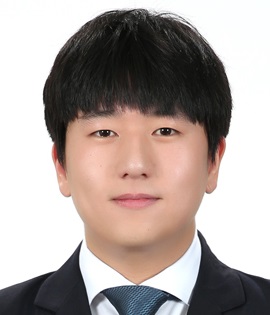}}]{Jee Seok Yoon}
	received a BS degree in Computer Science and Engineering from Korea University, Seoul, South Korea, in 2018. He is pursuing a PhD with the Department of Brain and Cognitive Engineering at Korea University, Seoul, South Korea. 
    
    His research interests include explainable AI, computer vision, and representation learning. 
\end{IEEEbiography}

\begin{IEEEbiography}[{\includegraphics[width=1in,clip,keepaspectratio]{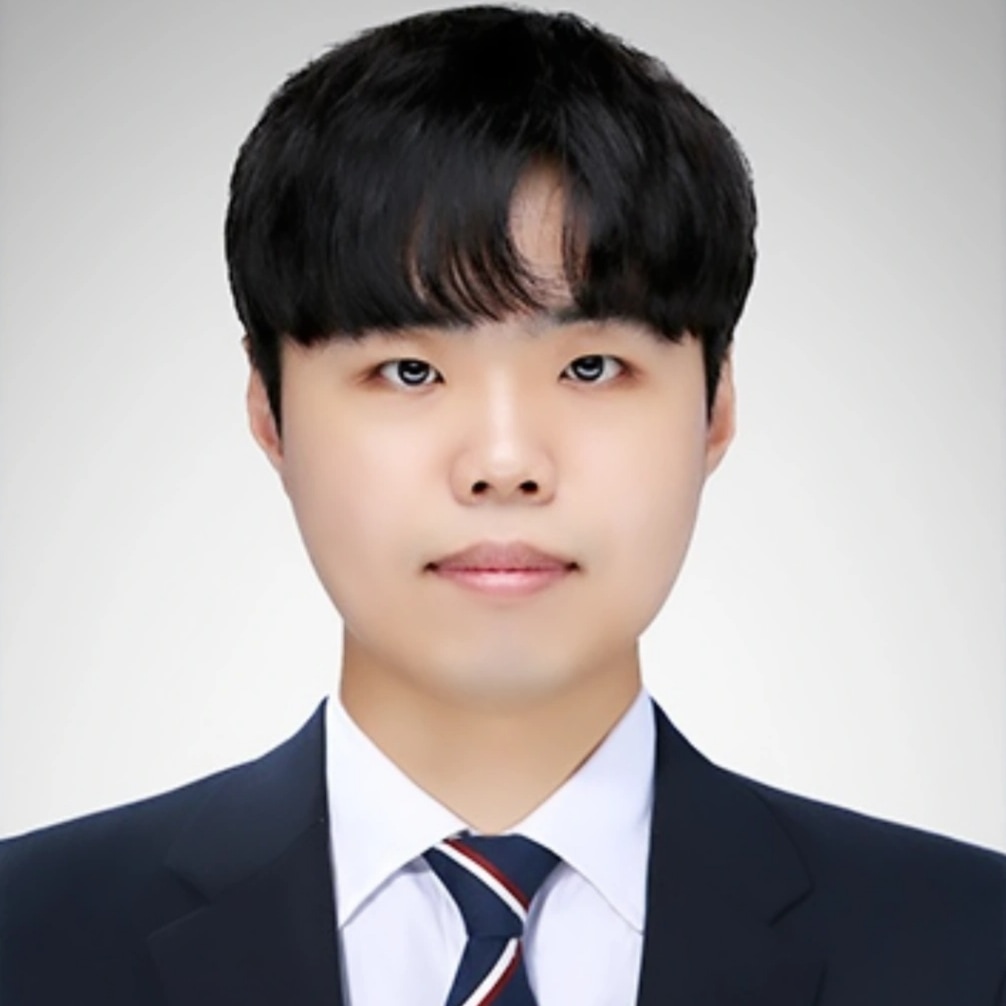}}]{Kwanseok Oh}
	received the BS degree in Electronic Control and Engineering from Hanbat National University, Daejeon, South Korea, in 2020. He is currently pursuing a Ph.D. degree with the Department of Artificial Intelligence at Korea University, Seoul, South Korea.
    
    His current research interests include explainable AI, computer vision, and representation learning. 
\end{IEEEbiography}

\begin{IEEEbiography}[{\includegraphics[width=1in,clip,keepaspectratio]{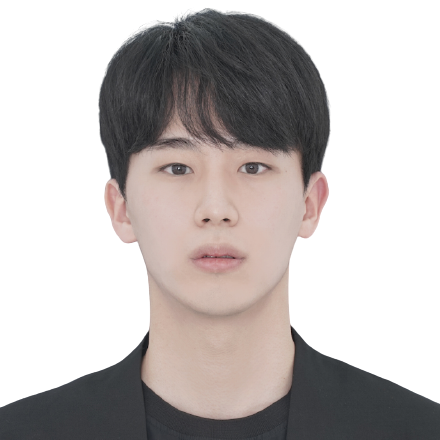}}]{Yooseung Shin}
	received the BS degree in Computer Science and Engineering from Hallym University, Chuncheon, South Korea, in 2022, and MS degree in Artificial Intelligence from Korea University, Seoul, South Korea, in 2024.

    His research interests include transfer learning, computer vision, and machine/deep learning.
\end{IEEEbiography}

\begin{IEEEbiography}[{\includegraphics[width=1in,clip,keepaspectratio]{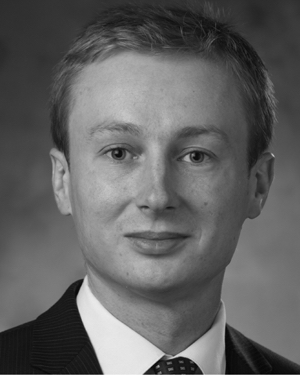}}]{Maciej A. Mazurowski}
	is currently an associate professor of radiology and electrical and computer engineering at Duke University. He is also a faculty with the Duke Medical Physics Program and a member of Duke Cancer Institute. His main research interest includes applications of machine learning including deep learning and statistical modeling, as well as computer vision algorithms to medicine. The particular focus in terms of applications is medical imaging in the context of cancer treatment, as well as understanding image interpretation process and error-making in radiology. 
\end{IEEEbiography}

\begin{IEEEbiography}[{\includegraphics[width=1in,clip,keepaspectratio]{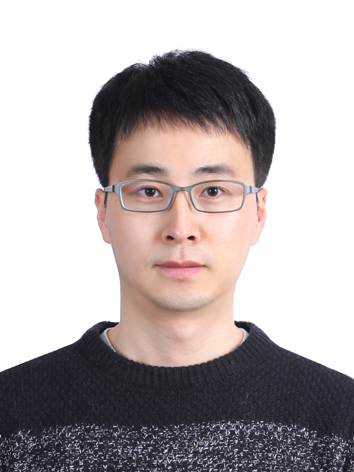}}]{Heung-Il Suk}
	is currently a Professor at the Department of Artificial Intelligence and an Adjunct Professor at the Department of Brain and Cognitive Engineering at Korea University. He was a Visiting Professor at the Department of Radiology at Duke University between 2022 and 2023. 
 
    He was awarded a Kakao Faculty Fellowship from Kakao and a Young Researcher Award from the Korean Society for Human Brain Mapping (KHBM) in 2018 and 2019, respectively. His research interests include causal machine/deep learning, explainable AI, biomedical data analysis, and brain-computer interface.

    Dr. Suk serves as an Editorial Board Member for Clinical and Molecular Hepatology (Artificial Intelligence Sector), Electronics, Frontiers in Neuroscience, Frontiers in Radiology (Artificial Intelligence in Radiology), International Journal of Imaging Systems and Technology (IJIST), and a Program Committee or a Reviewer for NeurIPS, ICML, ICLR, AAAI, IJCAI, CVPR, MICCAI, AISTATS, \etc
\end{IEEEbiography}

%% file: sections/appendix.tex
\setcounter{page}{0}

\onecolumn
\begin{appendices}
\begin{center}\LARGE\bfseries
Supplementary Material
\end{center}
\subsection{Benchmark Datasets}\label{sec:benchmark}
The field of domain generalization for \MIA relies primarily on {\JS customized datasets which are created by combining several private and publicly available datasets.
Although these datasets have greatly contributed to the field, there is a strong need to expand the spectrum of benchmark datasets.
In this section, we introduce three benchmark datasets for the domain generalization task in \MIA: Camelyon17-WILDS~\cite{bandi2018detection}, the M\&Ms challenge dataset~\cite{campello2021multi}, and the MIDOG challenge dataset~\cite{aubreville2023mitosis}.
To the best of our knowledge, these three are the only publicly open benchmark dataset for DG in \MIA.
For a comprehensive review of benchmark datasets for domain \textit{adaptation}, refer to~\cite{guan2021domain}.
}

\subsubsection{Existing benchmark datasets}
The \parggg{Camelyon17-WILDS}~\cite{bandi2018detection} dataset is derived from the Camelyon17 Challenge, which focuses on detecting metastasis in histopathological images of lymph nodes. This dataset, however, has been adapted specifically for domain generalization. It includes pathology images from two separate institutions and contains a total of 100 whole-slide images. The main challenge in this dataset comes from the inter-institutional variations, including differences in staining procedures and scanners used, which can significantly affect the performance of models.
The \parggg{M\&Ms challenge dataset}~\cite{campello2021multi} is a multi-center, multi-vendor, and multi-disease cardiac magnetic resonance (CMR) dataset. The M\&Ms challenge focuses on automatically segmenting the left ventricle, right ventricle, and myocardium in cardiac MR images, which are critical for diagnosing and managing various cardiovascular diseases. The dataset contains images from five different sites and five scanners, with patients suffering from five distinct pathologies. This heterogeneity poses a challenge for domain generalization, as models need to overcome variations in imaging protocols, equipment, and patient populations.
The \parggg{MIDOG challenge dataset}~\cite{aubreville2023mitosis} is focused on detecting mitotic figures in histopathological images, which is an important task for cancer diagnosis and grading. This dataset comprises images from five different hospitals, and the images have been digitized using different scanners and have undergone various staining procedures. These inter-hospital variations make it a challenging dataset for domain generalization.

\subsubsection{Emerging benchmark dataset}
One promising dataset for this purpose is the Brain Tumor Segmentation (BRATS) Challenge dataset. The BRATS dataset has been a valuable resource for neuroimaging researchers since its inception.
Since the 2022 challenge, the dataset has included additional cohorts from pediatric~\cite{kazerooni2023brain} and African~\cite{adewole2023brain} populations. These additions significantly increase the diversity of the dataset, making it an ideal resource for domain generalization research.
Including pediatric and African cohorts helps address two key areas of need in the field.
First, there is a growing acknowledgment of the need to ensure that machine learning models in healthcare are trained on diverse data representing various age groups.
The pediatric cohort in the BRATS dataset provides an opportunity to test and improve the performance of models in analyzing medical images from younger patients.
Second, the African cohort provides much-needed diversity in terms of ethnicity, helping to mitigate model biases and improve the generalizability of machine learning models across different ethnic groups.
Another emerging benchmark dataset for domain generalization is the Retinal OCT Fluid Challenge (RETOUCH)~\cite{bogunovic2019retouch}, a cross-site dataset of 70 OCT volumes with 3 sites.

\subsection{Publicly Open Datasets}
\input{tables/dataset.tex}

\end{appendices}

\twocolumn

%% file: tables/dataset.tex
\begin{scriptsize}
    
\begin{longtable}{ccccc}

\caption{List of datasets used by reviewed literature.}\\
    \toprule
     {\bfseries Task} & {\bfseries Organ} & {\bfseries Modality} & {\bfseries Abbreviation} & {\bfseries Ref.}\\
    \bottomrule
    \endfirsthead
    \toprule
     {\bfseries Task} & {\bfseries Organ} & {\bfseries Modality} & {\bfseries Abbreviation} & {\bfseries Ref.}\\
    \bottomrule
    \endhead

    \hline
    \multicolumn{5}{|r|}{{Continued on next page}} \\ \hline
    \endfoot
    
    \hline
    \multicolumn{5}{|r|}{{End of table}} \\ \hline
    \endlastfoot

    Classification & Brain & MRI & ADNI\footnote{Alzheimer’s Disease Neuroimaging Initiative, \url{https://adni.loni.usc.edu/}\label{f:adni}} & \cite{nguyen2023adversarially,wang2021harmonization} \\\cmruledata
 &  &  & iSTAGING\footnote{Imaging-based coordinate SysTem for AGing and NeurodeGenerative diseases consortium, \url{https://doi.org/10.1093/braincomms/fcac117}\label{f:istaging}} & \cite{wang2022embracing,wang2021harmonization} \\\cmruledataa
 &  & EEG & MAYO\footnote{Multicenter intracranial EEG dataset for classification of graphoelements and artifactual signals, \url{https://doi.org/10.6084/m9.figshare.c.4681208}\label{f:mayo}} & \cite{wang2022seeg} \\\cmruledata
 &  &  & FNUSA\footnote{Multicenter sEEG dataset, \url{https://www.kaggle.com/datasets/nejedlypetr/multicenter-intracranial-eeg-dataset}\label{f:fnusa}} & \cite{wang2022seeg} \\\cmruledataaa
 & Skin & Dermatology & ISIC\footnote{International Skin Imaging Collaboration, \url{https://api.isic-archive.com/collections/}\label{f:isic}} & \cite{li2020domain,goel2020model,bissoto2022artifact,nguyen2023lvm} \\\cmruledata
 &  &  & HAM10000\footnote{The Human Against Machine with 10000 training images, \url{https://doi.org/10.7910/DVN/DBW86T}\label{f:ham10000}} & \cite{li2020domain,bekkouch2021adversarial} \\\cmruledata
 &  &  & Fitzpatrick17K\footnote{Fitzpatrick17K, \url{https://github.com/mattgroh/fitzpatrick17k}\label{f:fitzpatrick17k}} & \cite{pakzad2022circle} \\\cmruledata
 &  &  & Dermofit\footnote{DERMOFIT dataset, \url{https://licensing.edinburgh-innovations.ed.ac.uk/product/dermofit-image-library}\label{f:dermofit}} & \cite{li2020domain} \\\cmruledata
 Classification & Skin & Dermatology & PH2\footnote{Ph2 Database, \url{https://www.fc.up.pt/addi/ph2\%20database.html}\label{f:ph2}} & \cite{li2020domain} \\\cmruledata
 &  &  & Derm7pt\footnote{7-point criteria evaluation Database, \url{https://derm.cs.sfu.ca/}\label{f:derm7pt}} & \cite{li2020domain} \\\cmruledataa
 &  & Histology & openLCH\footnote{Live Cell Histology, \url{https://doi.org/10.17867/10000161}\label{f:openlch}} & \cite{nguyen2023adversarially} \\\cmruledataaa
 & Colon & Histology & Kather16\footnote{Kather16, \url{https://doi.org/10.5281/zenodo.53169}\label{f:kather16}} & \cite{tomar2023tesla,scalbert2022test} \\\cmruledata
 &  &  & Kather18\footnote{Kather18, \url{https://doi.org/10.5281/zenodo.1214456}\label{f:kather18}} & \cite{tomar2023tesla,scalbert2022test,yamashita2021learning} \\\cmruledata
 &  &  & Kather19\footnote{Kather19, \url{https://doi.org/10.5281/zenodo.2530835}\label{f:kather19}} & \cite{yamashita2021learning} \\\cmruledata
 &  &  & CRC-TP\footnote{ColoRectal Cancer for Tissue Phenotyping, \url{https://warwick.ac.uk/fac/cross\_fac/tia/data/}\label{f:crc-tp}} & \cite{scalbert2022test} \\\cmruledata
 &  &  & AIDA-LNCO\footnote{Regional lymph node metastasis in colon adenocarcinoma dataset, \url{https://doi.org/10.23698/aida/lnco}\label{f:aida-lnco}} & \cite{stacke2020measuring} \\\cmruledataaa
 & Blood Cell & Histology & BCISC\footnote{Blood Cell Images for Segmentation and Classification dataset, \url{https://github.com/fpklipic/BCISC}\label{f:bcisc}} & \cite{chen2021d} \\\cmruledata
 &  &  & LISC\footnote{Leukocyte Images for Segmentation and Classification, \url{https://users.cecs.anu.edu.au/\textasciitilde{}hrezatofighi/Data/Leukocyte\%20Data.htm}\label{f:lisc}} & \cite{chen2021d} \\\cmruledata
 &  &  & BCCD\footnote{Blood Cell Count and Detection, \url{https://www.kaggle.com/datasets/paultimothymooney/blood-cells}\label{f:bccd}} & \cite{chen2021d} \\\cmruledataaa
 & Chest & X-ray & MIMIC-CXR\footnote{Medical Information Mart for Intensive Care-Chest X-Ray, \url{https://doi.org/10.13026/s5dg-6s42}\label{f:mimic-cxr}} & \cite{azizi2023robust,puli2021out,puli2022nuisances,zhang2022semi} \\\cmruledata
 &  &  & CXR8\footnote{ChestX-ray8, \url{https://nihcc.app.box.com/v/ChestXray-NIHCC}\label{f:cxr8}} & \cite{mahajan2021domain,zhang2022semi} \\\cmruledata
 &  &  & ChexPert\footnote{Chest eXpert (A Large Chest Radiograph Dataset), \url{https://stanfordmlgroup.github.io/competitions/chexpert/}\label{f:chexpert}} & \cite{azizi2023robust,mahajan2021connection,puli2021out,puli2022nuisances,mahajan2021domain,zhang2022semi} \\\cmruledata
 &  &  & RSNA-PD\footnote{RSNA Pneumonia Detection Challenge, \url{https://www.rsna.org/rsnai/ai-image-challenge/RSNA-Pneumonia-Detection-Challenge-2018}\label{f:rsna-pd}} & \cite{mahajan2021domain} \\\cmruledata
 &  &  & ChestX-ray14\footnote{NIH Chest X-rays, \url{https://www.kaggle.com/datasets/nih-chest-xrays/data}\label{f:chestx-ray14}} & \cite{azizi2023robust} \\\cmruledata
 &  &  & PadChest\footnote{PadChest: A Large Chest X-ray Image Dataset with Multi-Label Annotated Reports, \url{https://bimcv.cipf.es/bimcv-projects/padchest/}\label{f:padchest}} & \cite{zhang2022semi} \\\cmruledataaa
 & Breast & Histology & CAMELYON17\footnote{CAMELYON, \url{https://camelyon17.grand-challenge.org/}\label{f:camelyon17}} & \cite{scalbert2022test,chang2021stain,stacke2020measuring,wang2021variational,azizi2023robust,chen2023federated,gao2022out,yuan2022not,cheng2023adversarial} \\\cmruledataa
 &  & X-ray & DDSM\footnote{CBIS-DDSM: Breast Cancer Image Dataset, \url{https://wiki.cancerimagingarchive.net/x/lZNXAQ}\label{f:ddsm}} & \cite{wang2022domain,wang2023learning} \\\cmruledataaa
 & Knee & MRI & MOST\footnote{Multicenter Osteoarthritis Study, \url{https://agingresearchbiobank.nia.nih.gov/studies/most/}\label{f:most}} & \cite{karim2021deepkneeexplainer} \\\cmruledataaa
 & Lung & CT & LIDC-IDRI\footnote{The Lung Image Database Consortium and Image Database Resource Initiative, \url{https://doi.org/10.7937/K9/TCIA.2015.LO9QL9SX}\label{f:lidc-idri}} & \cite{wang2020proactive} \\\cmruledataaa
 & Retinal & OCT & A2ASDOCT\footnote{AREDS 2 Ancillary Spectral Domain Optical Coherence Tomography Study, \url{https://clinicaltrials.gov/study/NCT00734487}\label{f:a2asdoct}} & \cite{wang2020proactive} \\

    \midrule

    Segmentation & Brain & MRI & WMH\footnote{MICCAI White Matter Hyperintensity Challenge, \url{https://doi.org/10.34894/AECRSD}\label{f:wmh}} & \cite{zhao2022test} \\\cmruledata
 &  &  & ATLAS\footnote{Anatomical Tracings of Lesions After Stroke, \url{https://doi.org/10.3886/ICPSR36684}\label{f:atlas}} & \cite{yu2023san} \\\cmruledata
 &  &  & ABIDE\footnote{Autism Brain Imaging Data Exchange, \url{https://fcon\_1000.projects.nitrc.org/indi/abide/}\label{f:abide}} & \cite{karani2021test,10251660} \\\cmruledata
 &  &  & ISBI-MS\footnote{ISBI Longitudinal Multiple Sclerosis Lesion, \url{https://iacl.ece.jhu.edu/index.php/MSChallenge}\label{f:isbi-ms}} & \cite{kamraoui2022deeplesionbrain} \\\cmruledata
 &  &  & BraTS\footnote{Brain Tumor Segmentation Challenge, \url{https://www.med.upenn.edu/cbica/brats/}\label{f:brats}} & \cite{taleb2021multimodal,zhou2022generalizable,nguyen2023lvm,li2019privacy} \\\cmruledata
 &  &  & CC359\footnote{Calgary-Campinas 359, \url{https://www.ccdataset.com/download}\label{f:cc359}} & \cite{zakazov2022feather} \\\cmruledata
 Segmentation& Brain & MRI & MSSEG\footnote{MICCAI2016 MS Challenge Dataset, \url{https://portal.fli-iam.irisa.fr/msseg-challenge/}\label{f:msseg}} & \cite{kamraoui2022deeplesionbrain} \\\cmruledataa
 &  & fMRI, MRI & HCP\footnote{Human Connectome Project, \url{https://www.humanconnectome.org/}\label{f:hcp}} & \cite{karani2021test} \\\cmruledataaa
 & Retinal & Fundus & Drishti-GS\footnote{Retinal image dataset for optic disc and cup segmentation, \url{http://cvit.iiit.ac.in/projects/mip/drishti-gs/mip-dataset2/Home.php}\label{f:drishti-gs}} & \cite{liu2022single,lyu2022aadg,wang2020dofe,zhang2023domain,gu2022cddsa,zhang2022semi,zhang2022robust,gu2022contrastive} \\\cmruledata
 &  &  & STARE\footnote{Structured Analysis of the Retina dataset, \url{https://cecas.clemson.edu/\textasciitilde{}ahoover/stare/}\label{f:stare}} & \cite{liu2023ss,lyu2022aadg,wang2020dofe,hu2023map} \\\cmruledata
 &  &  & IDRiD\footnote{Indian Diabetic Retinopathy Image Dataset, \url{https://dx.doi.org/10.21227/H25W98}\label{f:idrid}} & \cite{lyu2022aadg,liu2021recursively,liu2022ordinal} \\\cmruledata
 &  &  & KDR\footnote{Kaggle Diabetic Retinopathy dataset, \url{https://kaggle.com/competitions/diabetic-retinopathy-detection}\label{f:kdr}} & \cite{lyu2022aadg,liu2021recursively,liu2022ordinal,li2023frequency,galappaththige2024generalizing} \\\cmruledata
 &  &  & RIM-ONE-r3\footnote{RIM-ONE Release 3, \url{https://medimrg.webs.ull.es/}\label{f:rim-one-r3}} & \cite{gunasinghe2022domain,liu2022single,lyu2022aadg,wang2020dofe,zhang2023domain,liu2021feddg,gu2022cddsa,zhang2022semi,zhang2022robust,gu2022contrastive} \\\cmruledata
 &  &  & DRHAGIS\footnote{DR HAGIS database, \url{https://personalpages.manchester.ac.uk/staff/niall.p.mcloughlin/}\label{f:drhagis}} & \cite{liu2023ss} \\\cmruledata
 &  &  & REFUGE\footnote{Retinal Fundus Glaucoma Challenge, \url{https://refuge.grand-challenge.org/}\label{f:refuge}} & \cite{gunasinghe2022domain,xu2023fourier,liu2022single,lyu2022aadg,wang2020dofe,zhang2023domain,xu2023federated,liu2021feddg,gu2022cddsa,zhang2022semi,zhang2022robust,gu2022contrastive} \\\cmruledata
 &  &  & CHASE\footnote{Child Heart and Health Study in England dataset, \url{https://researchdata.kingston.ac.uk/id/eprint/96}\label{f:chase}} & \cite{liu2023ss,lyu2022aadg,wang2020dofe} \\\cmruledata
 &  &  & E-Ophtha\footnote{e-ophtha database, \url{https://www.adcis.net/en/third-party/e-ophtha/}\label{f:e-ophtha}} & \cite{lyu2022aadg} \\\cmruledata
 &  &  & ARIA\footnote{Automated Retinal Image Analysis dataset, \url{https://www.researchgate.net/post/How\_can\_I\_find\_the\_ARIA\_Automatic\_Retinal\_Image\_Analysis\_Dataset/5964f84aed99e15c3140b3e6/citation/download}\label{f:aria}} & \cite{liu2023ss,hu2023map} \\\cmruledata
 &  &  & IOSTAR\footnote{IOSTAR vessel segmentation dataset, \url{https://www.retinacheck.org/download-iostar-retinal-vessel-segmentation-dataset}\label{f:iostar}} & \cite{liu2023ss,li2023frequency} \\\cmruledata
 &  &  & HRF\footnote{High-Resolution Fundus Image Database, \url{https://www5.cs.fau.de/research/data/fundus-images/}\label{f:hrf}} & \cite{lyu2022aadg,wang2020dofe} \\\cmruledata
 &  &  & LES-AV\footnote{LES-AV dataset, \url{https://doi.org/10.6084/m9.figshare.11857698}\label{f:les-av}} & \cite{li2023frequency} \\\cmruledata
 &  &  & PRIME-FP20\footnote{PRIME-FP20: Ultra-widefield Fundus Photography Vessel Segmentation Dataset, \url{https://doi.org/10.21227/ctgj-1367}\label{f:prime-fp20}} & \cite{hu2023map} \\\cmruledata
 &  &  & RIGA+\footnote{RIGA+ Dataset for Unsupervised Domain Adaptation in Medical Image Segmentation, \url{https://doi.org/10.5281/zenodo.6325549}\label{f:riga+}} & \cite{hu2023devil} \\\cmruledata
 &  &  & APTOS\footnote{APTOS 2019 Blindness Detection, \url{https://www.kaggle.com/c/aptos2019-blindness-detection}\label{f:aptos}} & \cite{galappaththige2024generalizing} \\\cmruledata
 &  &   & MESSIDOR\footnote{Feedback on a Publicly Distributed Image Database: The Messidor Database, \url{https://www.adcis.net/en/third-party/messidor/}\label{f:messidor}} & \cite{galappaththige2024generalizing} \\\cmruledataa
 &  & OCT & OCTA-500\footnote{A Retinal Dataset for Optical Coherence Tomography Angiography, \url{http://ieee-dataport.org/1951}\label{f:octa-500}} & \cite{hu2022domain,lyu2022aadg,hu2023map} \\\cmruledata
 &  &  & ROSE\footnote{Retinal OCTA SEgmentation dataset, \url{https://imed.nimte.ac.cn/dataofrose.html}\label{f:rose}} & \cite{hu2022domain,lyu2022aadg,hu2023map} \\\cmruledataa
 &  & FA & RECOVERY-FA19\footnote{RECOVERY-FA19: Ultra-widefield Fluorescein Angiography Vessel Detection Dataset, \url{https://doi.org/10.21227/m9yw-xs04}\label{f:recovery-fa19}} & \cite{hu2023map} \\\cmruledataaa
 & Prostate & MRI & NCI-ISBI\footnote{NCI-ISBI 2013 Challenge, \url{http://dx.doi.org/10.7937/K9/TCIA.2015.zF0vlOPv}\label{f:nci-isbi}} & \cite{karani2021test,liu2022single,tomar2023tesla,xu2022adversarial,chen2022maxstyle,zhou2023fedfa,liu2020shape,zhang2023domain,santhirasekaram2023topology,liu2021feddg,gao2023bayeseg,ouyang2022causality,gao2023desam} \\\cmruledata
 &  &  & I2CVB\footnote{Initiative for Collaborative Computer Vision Benchmarking, \url{http://i2cvb.github.io/}\label{f:i2cvb}} & \cite{liu2022single,tomar2023tesla,xu2022adversarial,chen2022maxstyle,zhou2023fedfa,liu2020shape,zhang2023domain,gao2023bayeseg,ouyang2022causality,gao2023desam} \\\cmruledata
 &  &  & PROMISE\footnote{Prostate MR Image Segmentation 2012, \url{https://promise12.grand-challenge.org/}\label{f:promise}} & \cite{karani2021test,liu2022single,xu2022adversarial,chen2022maxstyle,zhou2023fedfa,liu2020shape,zhang2023domain,liu2021feddg,gao2023bayeseg,ouyang2022causality,gao2023desam} \\\cmruledataaa
 & Prostate, Spinal & MRI & SCGM\footnote{Spinal Cord Grey Matter Segmentation Challenge, \url{http://cmictig.cs.ucl.ac.uk/niftyweb/challenge/}\label{f:scgm}} & \cite{tomar2023tesla,li2020domain,wang2021embracing,liu2021semi} \\\cmruledataaa
  & Spinal & CT & CSI\footnote{A multi-center milestone study of clinical vertebral CT segmentation, \url{http://spineweb.digitalimaginggroup.ca/}\label{f:csi}} & \cite{khandelwal2020domain} \\\cmruledata
    &  &  & xVertSeg\footnote{, \url{https://doi.org/10.17605/OSF.IO/NQJYW}\label{f:xvertseg}} & \cite{khandelwal2020domain} \\\cmruledata
 Segmentation & Spinal & CT & VerSe\footnote{VerSe: Large Scale Vertebrae Segmentation Challenge, \url{https://github.com/anjany/verse}\label{f:verse}} & \cite{khandelwal2020domain} \\\cmruledataaa
 & Cardiac & MRI & EMIDEC\footnote{Automatic Evaluation of Myocardial Infarction from Delayed-Enhancement Cardiac MRI, \url{https://emidec.com/}\label{f:emidec}} & \cite{gao2023bayeseg} \\\cmruledata
 &  &  & RVSC\footnote{Right Ventricle Segmentation Challenge, \url{https://rvsc.projets.litislab.fr/}\label{f:rvsc}} & \cite{karani2021test} \\\cmruledata
 &  &  & ACDC\footnote{Automated Cardiac Diagnosis Challenge, \url{https://www.creatis.insa-lyon.fr/Challenge/acdc/}\label{f:acdc}} & \cite{karani2021test,chen2022maxstyle,chen2020realistic,gao2023bayeseg} \\\cmruledata
 &  &  & M\&Ms\footnote{Multi-Centre, Multi-Vendor \& Multi-Disease Cardiac Image Segmentation Challenge 2020, \url{https://www.ub.edu/mnms/}\label{f:mms}} & \cite{ma2021histogram,li2021random,islam2022frequency,chen2022maxstyle,liu2021semi,santhirasekaram2023topology} \\\cmruledata
 &  &  & MS-CMRSeg\footnote{Multi-sequence Cardiac MR Segmentation Challenge, \url{https://zmiclab.github.io/zxh/0/mscmrseg19/}\label{f:ms-cmrseg}} & \cite{xu2022adversarial,su2023rethinking,chen2022maxstyle,ouyang2020self,gao2023bayeseg,ouyang2022causality} \\\cmruledataa
 &  & CT, MRI & MM-WHS\footnote{Multi-Modality Whole Heart Segmentation, \url{https://zmiclab.github.io/zxh/0/mmwhs/}\label{f:mm-whs}} & \cite{zhou2022generalizable,nguyen2023lvm,gao2023bayeseg} \\\cmruledataaa
 & Carotid & Ultrasound & SPLab\footnote{Signal processing laboratory, Brno University of Technology, \url{http://splab.cz/en/research/zpracovani-medicinskych-signalu/databaze/artery}\label{f:splab}} & \cite{bi2023mi} \\\cmruledataaa
 & Atrial & MRI & cDEMRIS\footnote{Cardiac Delayed Enhancement Segmentation Challenge, \url{https://figshare.com/articles/dataset/4214532}\label{f:cdemris}} & \cite{li2021atrialgeneral} \\\cmruledata
 &  &  & ASC\footnote{Atrial Segmentation Challenge, \url{https://www.cardiacatlas.org/atriaseg2018-challenge/}\label{f:asc}} & \cite{li2021atrialgeneral} \\\cmruledataaa
 & Abdominal & CT & SABSCT\footnote{MICCAI 2015 Multi-Atlas Abdomen Labeling Challenge, \url{https://doi.org/10.7303/syn3193805}\label{f:sabsct}} & \cite{xu2022adversarial,su2023rethinking,zhou2022generalizable,ouyang2020self,santhirasekaram2023topology,ouyang2022causality,chen2023ma} \\\cmruledata
 &  &  & LiTS\footnote{LiTS - Liver Tumor Segmentation Challenge, \url{http://www.lits-challenge.com/}\label{f:lits}} & \cite{wen2024denoising} \\\cmruledataa
  &  & CT, MRI & CHAOS\footnote{Combined (CT-MR) Healthy Abdominal Organ Segmentation challenge, \url{https://chaos.grand-challenge.org/}\label{f:chaos}} & \cite{taleb2021multimodal,xu2022adversarial,su2023rethinking,zhou2022generalizable,ouyang2020self,ouyang2022causality} \\\cmruledata
  &  & & MSD\footnote{The Medical Segmentation Decathlon, \url{http://medicaldecathlon.com/}\label{f:msd}} & \cite{chen2023ma} \\\cmruledataaa
 & Breast & Ultrasound & BUID\footnote{Breast Ultrasound Images Dataset, \url{https://www.kaggle.com/datasets/aryashah2k/breast-ultrasound-images-dataset}\label{f:buid}} & \cite{nguyen2023lvm} \\\cmruledataaa
 & Lung & X-ray & JSRT\footnote{Japanese Society of Radiological Technology, \url{http://db.jsrt.or.jp/eng.php}\label{f:jsrt}} & \cite{nguyen2023lvm} \\\cmruledataaa
 & Surgical Scene & Video Frames & EndoVis-Robot\footnote{Robotic Scene Segmentation Sub-Challenge, \url{https://endovissub2018-roboticscenesegmentation.grand-challenge.org/}\label{f:endovis-robot}} & \cite{seenivasan2022biomimetic,chen2023ma} \\\cmruledataaa
 & Gastrointestinal & Endoscopy & KvaSir\footnote{A Multi-Class Image-Dataset for Computer Aided Gastrointestinal Disease Detection, \url{https://datasets.simula.no/kvasir/}\label{f:kvasir}} & \cite{nguyen2023lvm} \\

\midrule

 Detection & Brain & EEG & CHB-MIT\footnote{Children's Hospital Boston (CHB)-MIT dataset, \url{https://doi.org/10.13026/C2K01R}\label{f:chb-mit}} & \cite{ayodele2020supervised} \\\cmruledata
 &  &  & TUSZ\footnote{Temple University Hospital (TUH) EEG Seizure Corpus dataset, \url{https://isip.piconepress.com/projects/tuh\_eeg/html/downloads.shtml}\label{f:tusz}} & \cite{ayodele2020supervised} \\\cmruledataaa
 & Chest & X-ray & COVID-QU-ex\footnote{COVID-QU-ex, \url{https://doi.org/10.34740/kaggle/dsv/3122958}\label{f:covid-qu-ex}} & \cite{muller2022shortcut} \\\cmruledata
 &  &  & BIMCV\footnote{A large annotated dataset of RX and CT images of COVID19 patients, \url{https://bimcv.cipf.es/bimcv-projects/bimcv-covid19/}\label{f:bimcv}} & \cite{yang2022full,zheng2023single} \\\cmruledata
 &  &  & Hannover-CV\footnote{COVID-19 Image Repository, \url{https://doi.org/10.25835/0090041}\label{f:hannover-cv}} & \cite{yang2022full} \\\cmruledata
 &  &  & COVID-CT\footnote{A CT Scan Dataset about COVID-19, \url{https://github.com/UCSD-AI4H/COVID-CT}\label{f:covid-ct}} & \cite{yang2022full} \\\cmruledata
 &  &  & ActualMed\footnote{Actualmed COVID-19 Chest X-ray Dataset Initiative, \url{https://github.com/agchung/Actualmed-COVID-chestxray-dataset}\label{f:actualmed}} & \cite{yang2022full} \\\cmruledata
 &  &  & RSNA-BA\footnote{RSNA Pediatric Bone Age Challenge (2017), \url{https://www.rsna.org/rsnai/ai-image-challenge/rsna-pediatric-bone-age-challenge-2017}\label{f:rsna-ba}} & \cite{yang2022full} \\\cmruledata
 &  &  & VinDr\footnote{VinDr-CXR: An open dataset of chest X-rays with radiologist’s annotations, \url{https://vindr.ai/datasets/cxr}\label{f:vindr}} & \cite{nguyen2023lvm} \\\cmruledataaa
 & Breast & Histology & TUPAC\footnote{Tumor Proliferation Assessment Challenge, \url{https://tupac.grand-challenge.org}\label{f:tupac}} & \cite{xu2022improved,fernandez2024uninformed} \\\cmruledataa
 Detection & Breast & X-ray & INbreast\footnote{INbreast: toward a full-field digital mammographic database, \url{https://www.kaggle.com/datasets/ramanathansp20/inbreast-dataset}\label{f:inbreast}} & \cite{li2021domain} \\\cmruledataaa
 & Tissue & Histology & MIDOG\footnote{The Mitosis Domain Generalization Challenge, \url{https://imig.science/midog/}\label{f:midog}} & \cite{kondo2022tackling,fernandez2024uninformed} \\\cmruledata
 &  &  & MITOS-ATYPIA-14\footnote{MITOS-ATYPIA 2014 challenge of ICPR, \url{https://mitos-atypia-14.grand-challenge.org/Dataset/}\label{f:mitos}} & \cite{fernandez2024uninformed} \\
 &  &  & CCMCT\footnote{A large-scale dataset for mitotic figure assessment on whole slide images of canine cutaneous mast cell tumor, \url{https://github.com/DeepMicroscopy/MITOS_WSI_CCMCT}\label{f:ccmct}} & \cite{fernandez2024uninformed} \\

\midrule

Restoration & Retinal & Fundus & DRIVE\footnote{Digital Retinal Images for Vessel Extraction, \url{https://drive.grand-challenge.org/}\label{f:drive}} & \cite{liu2022domain,hu2022domain,liu2023ss,lyu2022aadg,wang2020dofe,nguyen2023lvm,li2023frequency,hu2023map} \\\cmruledata
 &  &  & Kag-Cat\footnote{Kaggle cataract dataset, \url{https://www.kaggle.com/datasets/jr2ngb/cataractdataset}\label{f:kag-cat}} & \cite{liu2022domain} \\

\midrule

Localization & Surgical Scene & Video Frames & SurgicalActions160\footnote{The ITEC SurgicalActions160 Dataset, \url{https://ftp.itec.aau.at/datasets/SurgicalActions160/}\label{f:surgicalactions160}} & \cite{philipp2022dynamic} \\\cmruledata
 &  &  & Cataract-101\footnote{Cataract-101 Video Dataset, \url{https://ftp.itec.aau.at/datasets/ovid/cat-101/}\label{f:cataract-101}} & \cite{philipp2022dynamic} \\

\midrule

Reconstruction & Blood & Histology & MalariaScreener\footnote{Malaria Screener, \url{https://lhncbc.nlm.nih.gov/LHC-downloads/downloads.html\#malaria-datasets}\label{f:malariascreener}} & \cite{ilse2020diva} \\\cmruledataaa
 & Brain & MRI & BrainWeb\footnote{BrainWeb, \url{https://brainweb.bic.mni.mcgill.ca/brainweb/}\label{f:brainweb}} & \cite{song2019coupled} \\
\bottomrule

\end{longtable}
\end{scriptsize}